\documentclass[fleqn,usenatbib]{mnras}

\usepackage{newtxtext,newtxmath}

\usepackage[T1]{fontenc}

\DeclareRobustCommand{\VAN}[3]{#2}
\let\VANthebibliography\thebibliography
\def\thebibliography{\DeclareRobustCommand{\VAN}[3]{##3}\VANthebibliography}

\usepackage{graphicx}	
\usepackage{amsmath}	
\usepackage{orcidlink}

\title[Robustness of the Auriga model]{Quantifying the intrinsic variability due to randomness of the Auriga galaxy formation model}

\author[R. Pakmor et al.]{R\"udiger Pakmor$^1$\thanks{rpakmor@mpa-garching.mpg.de}\orcidlink{0000-0003-3308-2420},
Rebekka Bieri$^{2}$\orcidlink{0000-0002-4554-4488},
Francesca Fragkoudi$^{3}$\orcidlink{0000-0002-0897-3013},
Facundo A.~G\'{o}mez,$^{4}$\orcidlink{0000-0003-4232-8584},
\newauthor
Robert J.~J.~Grand$^{5}$\orcidlink{0000-0001-9667-1340},
Christine M. Simpson$^{6}$\orcidlink{0000-0001-9985-1814},
Rosie Y. Talbot$^1$\orcidlink{0000-0001-9393-7879},
Freeke van~de~Voort$^7$\orcidlink{0000-0002-6301-638X},
\newauthor
Maria Werhahn$^1$\orcidlink{0000-0003-4984-4389}
\vspace*{0.1cm}\\
$^{1}$Max-Planck-Institut f\"{u}r Astrophysik, Karl-Schwarzschild-Str. 1, D-85748, Garching, Germany\\
$^{2}$Department of Astrophysics, University of Zurich, Zurich, Switzerland\\
$^{3}$Institute for Computational Cosmology, Department of Physics, Durham University, South Road, Durham DH1 3LE, UK \\
$^{4}$Departamento de Astronom\'{i}a, Universidad de La Serena, Av.~Juan Cisternas 1200 Norte, La Serena, Chile\\
$^{5}$Astrophysics Research Institute, Liverpool John Moores University, 146 Brownlow Hill, Liverpool, L3 5RF, UK\\
$^{6}$Argonne Leadership Computing Facility, Argonne National Laboratory, Lemont, IL 60439, USA\\
$^{7}$Cardiff Hub for Astrophysics Research and Technology, School of Physics and Astronomy, Cardiff University, Queen’s Buildings, Cardiff CF24 3AA, UK\\
}

\date{Accepted XXX. Received YYY; in original form ZZZ}

\pubyear{\the\year{}}

\begin{document}
\label{firstpage}
\pagerange{\pageref{firstpage}--\pageref{lastpage}}
\maketitle

\begin{abstract}
Numerical simulations have become an indispensable tool in astrophysics. To interpret their results, it is critical to understand their intrinsic variability, that is, how much the results change with numerical noise or inherent stochasticity of the physics model.
We present a set of seven realisations of high-resolution cosmological zoom-in simulations of a Milky Way-like galaxy with the Auriga galaxy formation model. All realisations share the same initial conditions and code parameters, but draw different random numbers for the inherently stochastic parts of the model.
We show that global galaxy properties at $z=0$, including stellar mass, star formation history, masses of stellar bulge and stellar disc, the radius and height of the stellar disk change by less than $10\%$ between the different realisations, and that magnetic field structures in the disc and the halo are very similar.
In contrast, the star formation rate today can vary by a factor of two and the internal morphological structure of the stellar disc can change.
The time and orbit of satellite galaxies and their galaxy properties when falling into the main halo are again very similar, but their orbits start to deviate after first pericenter passage.
Finally, we show that changing the mass resolution of all matter components by a factor of $8$ in the Auriga model changes galaxy properties significantly more than the intrinsic variability of the model, and that these changes are systematic. This limits detailed comparisons between simulations at different numerical resolutions.
\end{abstract}




\section{Introduction}
Galaxies are complex systems governed by many different physical processes that are often strongly coupled. Numerical simulations are an indispensable tool to model these physical processes and their interactions in detail to obtain a full picture of the formation and evolution of galaxies. Cosmological simulations in particular are critical, because they allow us to self-consistently model galaxies in their full environment and over entire cosmic history \citep{Vogelsberger2020,Crain2023}.

Recent cosmological galaxy simulations produce galaxy populations generally consistent with observations \citep{Vogelsberger2014Illustris,Schaye2005Eagle,Springel2018,Dave2019Simba,Dubois2016HorizonAGN,Pakmor2023MTNG}. Comparison with observations usually focuses on ensemble averages between a population of simulated galaxies and a population of observed galaxies. Ideally, this averages out galaxy to galaxy variability from selection effects. Moreover, it also reduces the impact of variability intrinsic to the simulations. However, the robustness of properties of simulated galaxies is hard to judge for individual galaxies, and the intrinsic variability is likely significant for many galaxy formation models \citep{Keller2019,Genel2019,Borrow2023}.

Quantifying the impact of specific physical processes from simulations of a large number of galaxies at reasonable resolution is computationally unfeasible, because every single box is expensive already \citep{Nelson2019TNGPublic, VillaescusaNavarro2021}. Therefore, galaxy simulations focusing on the study of physical processes by comparing simulations run with different physics models or parameters usually only simulate a small number of galaxies with a zoom-in approach. This way they focus essentially all resolution elements in one main galaxy of interest. They then simulate the same galaxy multiple times with different physical models, for example to understand the impact of magnetic fields on galaxies \citep{Pakmor2017, Whittingham2021,Whittingham2023} and the circum-galactic medium \citep{vandeVoort2021}, cosmic rays \citep{Buck2019, Hopkins2019, MartinAlvarez2023b, Montero2024, Bieri2025}, different supermassive black hole models \citep[see, e.g.][]{Irodotou2022}, or the impact of systematic variations of the assembly history of a galaxy \citep{Davies2021,Davies2022,Rey2023,Joshi2024}. Then, however, it becomes critical to understand the robustness of the simulations for individual galaxies.

A fundamental source of variability for essentially all numerical simulations is that floating-point operations do not commute. Rather, the exact result will depend on the order of operations, for example the order in which various partial forces are added up to a total force. \citet{Genel2019} showed that while these differences are initially tiny, they can grow quickly for sufficiently chaotic systems. Moreover, they demonstrated that even though a tiny change in the simulation in one place will not change the properties of a large population of galaxies very much, individual galaxies can eventually look very different.

A much more important source of variability for galaxy formation simulations, that also significantly amplifies variability from round-off errors, are stochastic operations built into the model, for example in the models for star formation of stellar feedback. Because galaxy formation includes many physical processes that are strongly coupled, any local deviation, for example because a different set of random numbers is drawn, will quickly propagate. Any change of the random number seed (that changes the set of random numbers used) or round-off errors (for example because we use a different number of tasks or a different machine to run on) will produce equally valid results for a given code and model. We can therefore use a set of simulations of the same initial conditions with the same code and model parameters, but different random number seeds, to quantify the intrinsic variability of a simulation. 

Despite the fundamental importance for galaxy physics studies, little work has been done focusing on the intrinsic variability of galaxy simulations. \citet{Keller2019} studied the intrinsic variance in simulations of isolated dwarf galaxies and cosmological simulations of Milky Way-like galaxies. Using different feedback models they ran pairs of dwarf galaxies with the same model and re-simulated a dwarf galaxy with one model $128$ times. They found a typical relative variation in stellar mass after $1\,\mathrm{Gyr}$ of $5\%$ to $10\%$, with differences up to a factor of $2$ in extreme cases. They also evolved pairs of Milky Way-like galaxies with different feedback models in a fully cosmological setting to $z=0.9$. There they found significantly larger deviations in stellar mass compared to the isolated dwarf galaxies, with persisting differences in stellar mass larger than $10\%$. They also show that the galaxies can appear qualitatively different.

Focusing on the impact of systematic modifications to the initial conditions, \citet{Davies2021,Davies2022} ran $9$ realisations of one cosmological zoom simulation with different random number seeds for five sets of initial conditions to disentangle the intrinsic variability of the \textsc{eagle} galaxy formation model \citep{Schaye2005Eagle} from the effect of modifying the initial conditions. They found that changing the initial conditions and thereby the assembly history of a Milky Way-like galaxy is a significantly bigger effect than the intrinsic variability of the \textsc{eagle} model. They still found an intrinsic variation of the stellar mass at $z=0$ of up to a factor of $2$ for one set of initial conditions, and generally significant changes to the rotational support of the disk of the main galaxy at $z=0$.

Lastly, to quantify the variance from stochastic modelling of star formation and feedback in the SWIFT-Eagle model, \citet{Borrow2023} ran $16$ realisations of a $25\,\mathrm{Mpc}^3$ cosmological box. The \textsc{swift} code is not deterministic anyway as a result of its implementation of task-based parallelism that does not preserve the order or numerical operations. 
That is, if the simulation is run again on the same machine with the same number of tasks, it produces slightly different results. Therefore, they did not need to change the random number seed, because any initial deviation quickly changes the specific random number used for future operations. They found that the stellar mass even of well resolved individual galaxies can vary by $25\%$ in the \textsc{SWIFT-eagle} model.

In this paper we analyse the intrinsic variance of the Auriga galaxy formation model \citep{Auriga}. Star formation histories and satellite mass functions of our set of $7$ cosmological zoom simulations of the same galaxy have already been shown in \citet{Grand2021} to understand and quantify resolution effects on the satellite population of a Milky Way-like galaxies. They found that the stochastic variability at fixed resolution for the Auriga model is significantly smaller than changes introduced by changing the numerical resolution, and that the satellite population is robust to stochastic variations as well. Here we present a more general and detailed quantitative analysis of the same $7$ simulations with a focus on the central galaxy at $z=0$.

\begin{figure*}
    \centering
    \includegraphics[width=\textwidth]{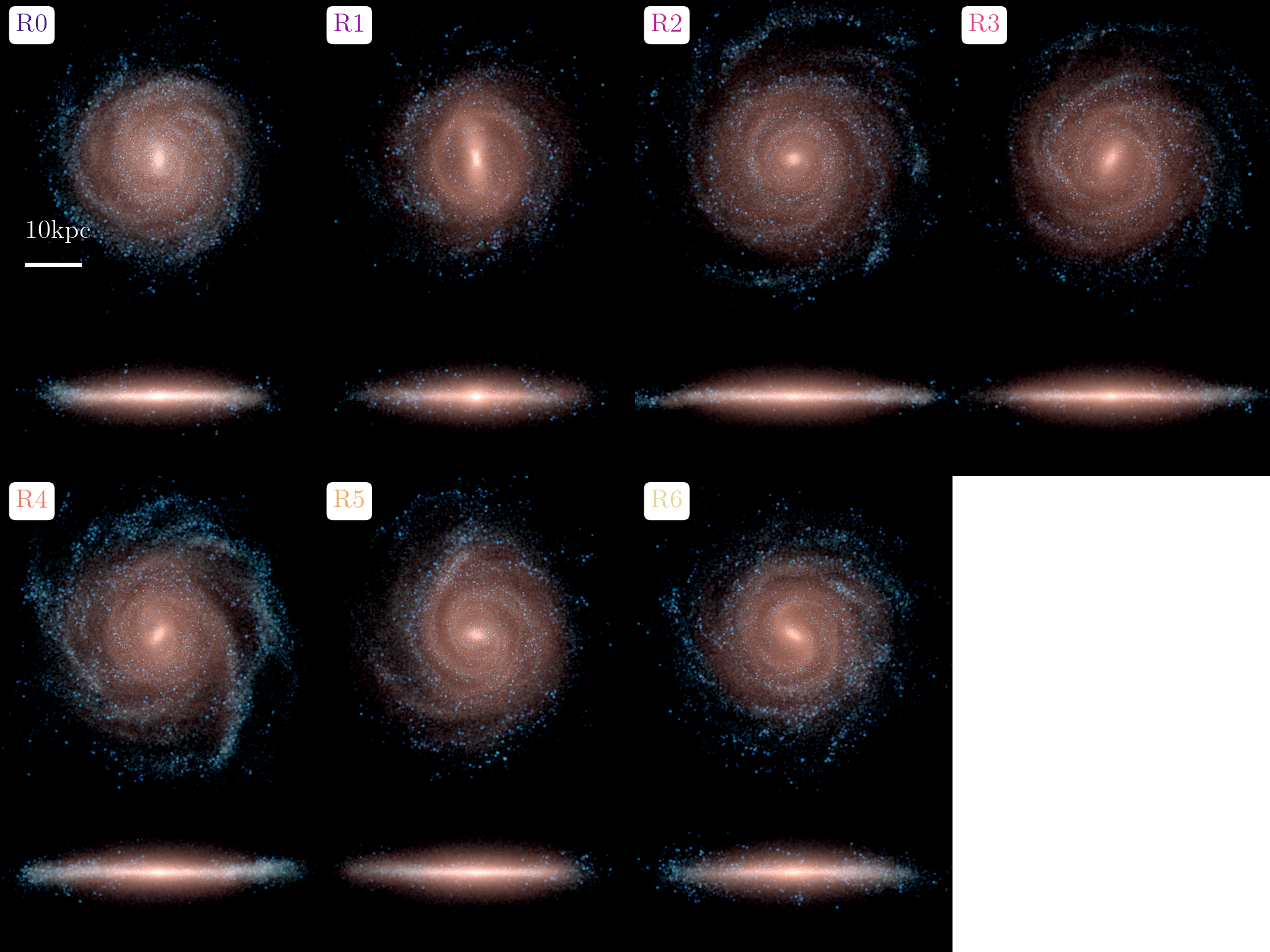}
    \caption{Stellar light projections of all seven realisations of Au-6 at resolution L4. All projections cover a $60\,\mathrm{kpc}\times60\,\mathrm{kpc}$ box for the face-on projections and a $60\,\mathrm{kpc}\times30\,\mathrm{kpc}$ for the edge-on projections. The galaxies all look very similar in colour and size. However, there is variation in their internal structure, in particular in the strength of the non-axisymmetric bar-like structure in the centre and their spiral features.}
    \label{fig:stellarlight}
\end{figure*}

\begin{table*}
\textbf{Realisations at standard resolution (L4)}
\begin{center}
\setlength\tabcolsep{3pt}
\begin{tabular}{ c c c c c c c c c c c c c }
\hline
Realisation & $M_{200\mathrm{c}}$ & $M_*$ & $M_{*,R>50\mathrm{kpc}}$ & $\dot{M}_{1\mathrm{Gyr}}$ & $V_\mathrm{c,max}$ & $M_\mathrm{disc}$ & $M_\mathrm{bulge}$ & $R_\mathrm{disc}$ & $H_\mathrm{disc}$ & $\sigma_\mathrm{r}$ & $\sigma_\mathrm{z}$ & $\left|L_*\right|$ \\
& $[10^{12}\rmn{M}_\odot]$ & $[10^{10}\rmn{M}_\odot]$ & $[10^{10}\rmn{M}_\odot]$ & $[\rmn{M}_\odot / \rmn{yr}]$ & $[\rmn{km/s}]$ & $[10^{10}\rmn{M}_\odot]$ & $[10^{10}\rmn{M}_\odot]$ & $[\rmn{kpc}]$ & $[\rmn{kpc}]$ & $[\rmn{km/s}]$ & $[\rmn{km/s}]$ & $[\rmn{10^{18}\rmn{M}_\odot\,kpc\,km/s}]$ \\
\hline
0 & $1.05$ & $5.66$ & $0.39$ & $3.7$ & $213$ & $0.42$ & $4.81$ & $3.9$ & $0.85$ & $49.6$ & $40.5$ & $6.5$\\
1 & $1.04$ & $5.47$ & $0.78$ & $1.6$ & $211$ & $0.42$ & $4.13$ & $4.1$ & $0.88$ & $50.3$ & $40.7$ & $5.7$\\
2 & $1.03$ & $5.09$ & $0.43$ & $2.1$ & $200$ & $0.42$ & $4.18$ & $5.2$ & $0.93$ & $46.5$ & $38.2$ & $6.8$\\
3 & $1.03$ & $5.32$ & $0.46$ & $1.6$ & $203$ & $0.42$ & $4.33$ & $4.8$ & $0.88$ & $45.0$ & $37.8$ & $6.9$\\
4 & $1.05$ & $5.39$ & $0.72$ & $3.3$ & $206$ & $0.44$ & $4.22$ & $4.7$ & $0.89$ & $47.7$ & $39.8$ & $6.4$\\
5 & $1.04$ & $5.42$ & $0.41$ & $2.4$ & $206$ & $0.43$ & $4.54$ & $4.7$ & $0.87$ & $46.6$ & $38.5$ & $6.9$\\
6 & $1.05$ & $5.36$ & $0.46$ & $2.0$ & $209$ & $0.39$ & $4.45$ & $4.0$ & $0.91$ & $46.8$ & $40.5$ & $5.9$\\
\hline
mean & $1.04$ & $5.39$ & $0.52$ & $2.4$ & $207$ & $0.42$ & $4.38$ & $4.5$ & $0.89$ & $47.5$ & $39.4$ & $6.5$\\
$\sigma$ & $0.01$ & $0.17$ & $0.16$ & $0.8$ & $4.6$ & $0.02$ & $0.24$ & $0.5$ & $0.03$ & $1.9$ & $1.2$ & $0.5$\\
$\sigma$ / mean $[\%]$ & $1.0$ & $3.2$ & $30.6$ & $33.6$ & $2.2$ & $4.0$ & $5.5$ & $10.7$ & $3.1$ & $3.9$ & $3.1$ & $7.2$\\
\hline
\end{tabular}
\end{center}
\caption{Global properties of our galaxy in all $7$ realisations at $z=0$. The columns show from left to right the mass of the halo $M_{200\mathrm{c}}$, that is the mass within a sphere around the potential minimum of the halo where the mean density is $200\times$ the critical density of the universe, the stellar mass of the main subhalo $M_*$, the stellar mass in the halo at a distance larger than $50\,\mathrm{kpc}$ $M_{*,R>50\mathrm{kpc}}$, the star formation rate in the last Gyr $\dot{M}_{1\mathrm{Gyr}}$, the maximum circular velocity $V_\mathrm{c,max}$, the mass of the stellar disc $M_\mathrm{disc}$ and bulge $M_\mathrm{bulge}$, the scale radius $R_\mathrm{disc}$ and height $H_\mathrm{disc}$ of the stellar disc, the radial $\sigma_\mathrm{r}$ and vertical $\sigma_\mathrm{z}$ velocity dispersion in the stellar disc at a radius of $8\,\mathrm{kpc}$, and the total angular momentum of the stellar disc $\left|L_*\right|$. The last three rows show the mean value, variance, and the variance relative to the mean for all quantities over all $7$ realisations. Most properties vary by $\lesssim 10\%$. Notably the stellar mass of the halo and the star formation rate at $z=0$ vary by $\approx 30\%$.}
\label{tab:lvl4}
\end{table*}

\begin{figure*}
    \centering
    \includegraphics[width=\textwidth]{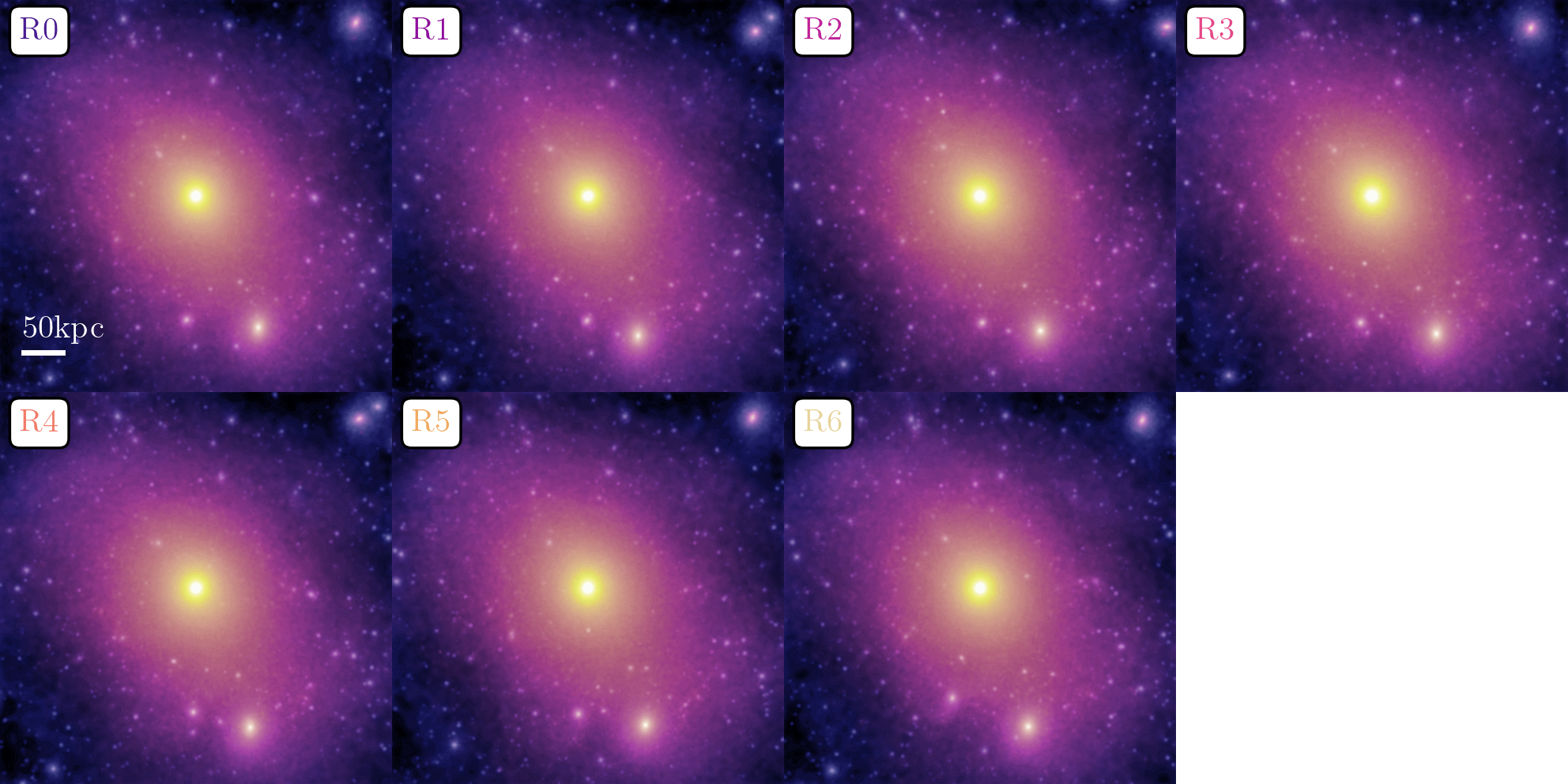}
    \caption{Projections of dark matter surface density and velocity dispersion with a two-dimensional colormap \citep{Springel2006}. All projections cover a $500\,\mathrm{kpc}\times500\,\mathrm{kpc}$ box. The panels show all seven realisations of Au-6 at resolution L4. The properties of the main dark matter halo as well as the positions of the most massive and many smaller satellites are very similar.}
    \label{fig:darkmatter}
\end{figure*}

\begin{figure}
    \centering
    \includegraphics[width=0.95\linewidth]{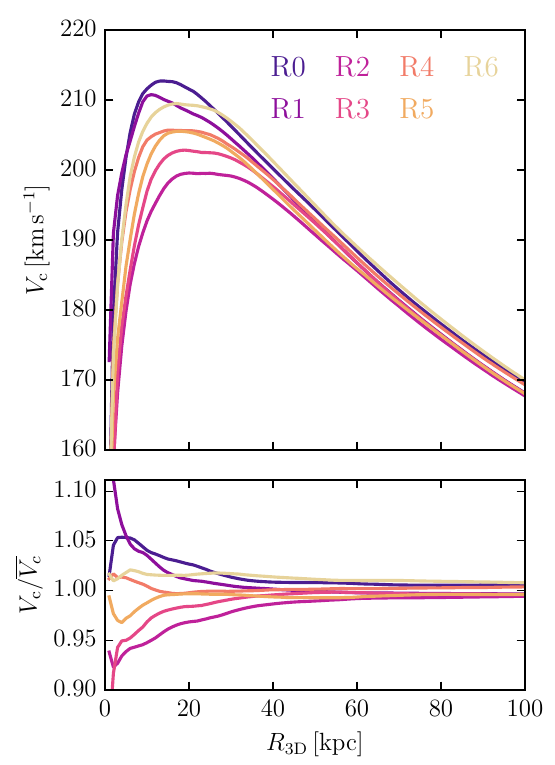}
    \caption{Circular velocity curves computed from the total enclosed mass of all seven realisations of Au-6 at resolution L4. The main difference is in the value of the peak or plateau around $20\,\mathrm{kpc}$. At larger radii circular velocity agrees on the per cent level.}
    \label{fig:rotationcurve}
\end{figure}

\section{Simulations}

Our simulations are re-simulations of the cosmological zoom-in simulation Au-6 of the Auriga project \citep{Auriga}. The Auriga halos are selected from the dark matter only simulation box of the \textsc{eagle} project \citep{Schaye2005Eagle}. They are chosen to have a halo mass of $10^{12}\,\mathrm{M_\odot} < R_{200,\mathrm{c}} < 2\times10^{12}\,\mathrm{M_\odot}$ at $z=0$ and also fulfil a mild isolation criterion. Each halo's particles are traced back to the initial conditions to determine its Lagrangian region at high-redshift. The Lagrangian region is resampled at higher resolution and augmented with an added safety buffer.  The halo is then re-simulated with the higher resolution in the Lagrangian region and the rest of the box is de-refined to a much lower resolution. In this way, the zoom-in simulation focuses almost all resolution elements on a single halo and its environment, but keeps the large-scale structure, and in particular tidal gravitational forces acting on the halo, in place.

The simulations are run with the moving-mesh magnetohydrodynamics code \textsc{arepo} \citep{Arepo,Pakmor2016,Weinberger2020}. It solves gravity with a combined tree and particle-mesh approach \citep{Springel2005b,Gadget4}. Gas is evolved on a moving Voronoi mesh with a second order finite volume scheme \citep{Pakmor2016}. The Auriga galaxy formation model adds the physical processes that are relevant for the formation and evolution of galaxies. Specifically, it includes radiative cooling in the form of primordial and metal line cooling \citep{Vogelsberger2013}, an effective model for the interstellar medium and the formation of stars \citep{Springel2003}, an effective model for galactic winds driven by stellar feedback, mass return from stars via stellar winds and supernovae, and a model for the seeding and growth of, as well as feedback from, supermassive black holes \citep{Auriga}.

\begin{figure}
    \centering
    \includegraphics[width=\linewidth]{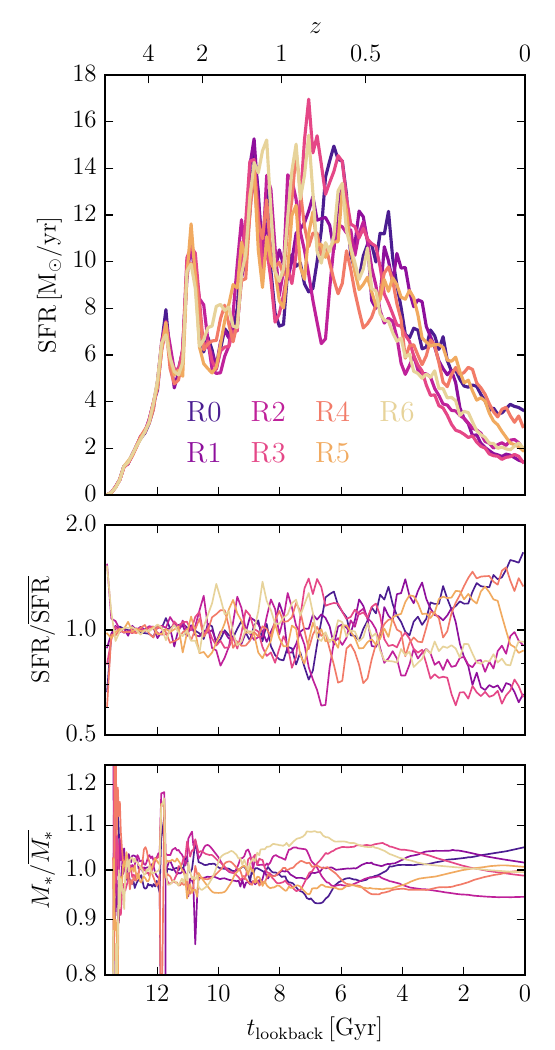}
    \caption{Star formation histories of all seven realisations of Au-6 at standard resolution L4 (top panel). They are essentially identical before $z=1$. As the star formation rates decline towards $z=0$, their relative differences (middle panel) increase, up to a deviation of $\sim~50\%$ relative to the mean. The total stellar mass deviates only on the level of a few percent over the whole evolution (bottom panel).}
    \label{fig:sfh}
\end{figure}

\begin{figure*}
    \centering
    \includegraphics[width=\textwidth]{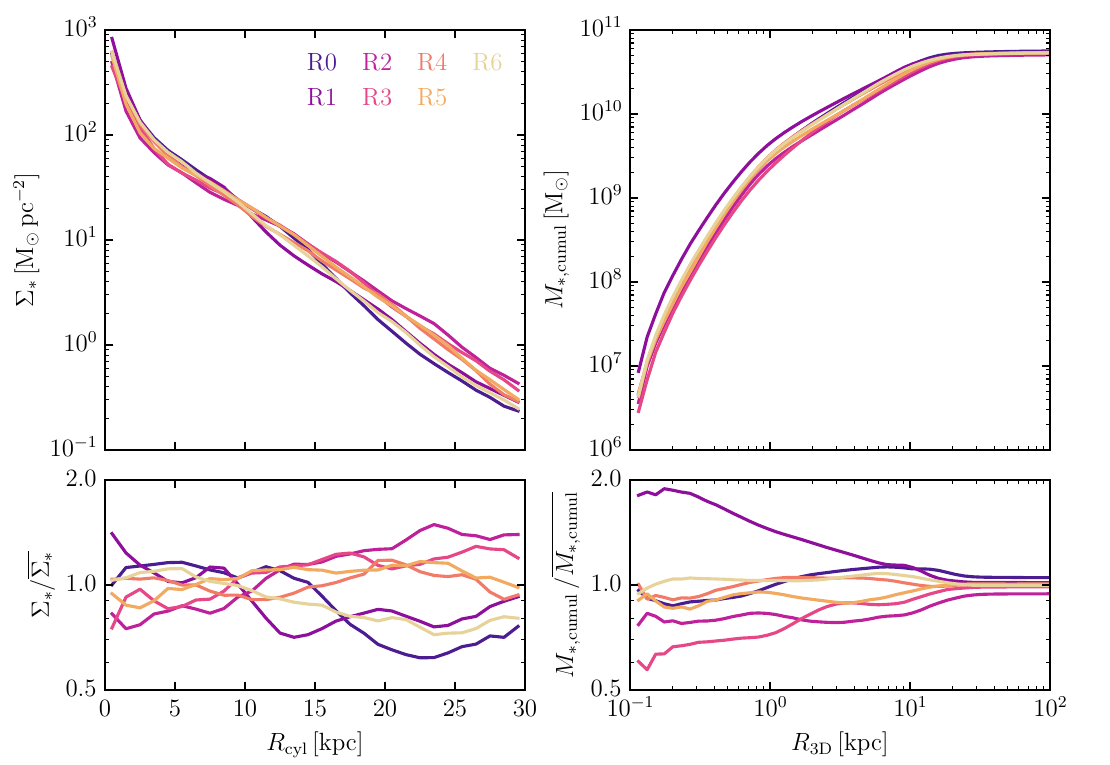}
    \caption{Stellar mass surface density (left panel) and cumulative stellar mass (right panel) for all seven realisations of Au-6 at resolution L4 The surface density profiles are very similar in the inner part as well as in their slope in the disc. The total stellar mass in the full halo is very similar as well, but the distribution in the halo varies somewhat between realisations.}
    \label{fig:stellar_density_profiles}
\end{figure*}

The main source of intrinsic randomness in the Auriga model is the formation of star particles and (galactic) wind particles. Both remove some or all gas from a cell (in the second case removing the cell completely) to create a collisionless particle. They are both realised stochastically, where the creation of a star particle models star formation and the wind particles are a model for the generation of galactic winds from stellar feedback. In each timestep, each star-forming cell has a non-zero probability to create a star particle or a wind particle \citep{Auriga}. To realise this probability,  a random number for each cell is drawn that determines whether the cell creates either particle or not. Once created, the wind particles have a second set of random numbers drawn to determine their initial direction of motion before they recouple to the gas mesh.

Additionally, the feedback of the supermassive black hole in the low accretion rate mode \citep{Auriga} randomly places hot bubbles in the circum-galactic medium, however, in the regime of the Auriga galaxies, this type of feedback is usually subdominant \citep{Auriga,Irodotou2022}.

In addition to randomness introduced by the physics model, we randomize the centre of the domain in every domain decomposition to decorrelate and therefore reduce force errors \citep{Gadget4}. The refinement also introduces randomness when deciding where to exactly introduce a new mesh-generating point when a cell is split \citep{Arepo}.

Importantly, the outcome of a simulation with the Auriga model (and with \textsc{arepo} in general) is deterministic. Therefore, rerunning the same simulation again on the same machine with the same parallel setup and libraries will lead to exactly identical results. To test the intrinsic variance of the Auriga model in a controlled way, we thus re-do one simulation seven times, with identical initial conditions, code, and model parameters, only changing the random number seed in \textsc{arepo} for every re-simulation. We call these re-simulations different realisations of the same simulation. In this way, all $7$ realisations are equally valid outcomes of the Auriga model for the given initial conditions, and the differences between the different realisations directly allow us to quantify the intrinsic variance of the model. For the resimulations we chose Au-6 of the Auriga project \citep{Auriga} at the standard resolution L4 (that is, with a baryonic mass resolution of $5\times10^4\,\mathrm{M_\odot}$ and a dark matter mass resolution of $3\times10^5\,\mathrm{M_\odot}$). We re-simulate Au-6 seven times, with identical initial conditions, code, and model parameters. We only change the random number seed in \textsc{arepo} for every simulation. 

\begin{figure*}
    \centering
    \includegraphics[width=\linewidth]{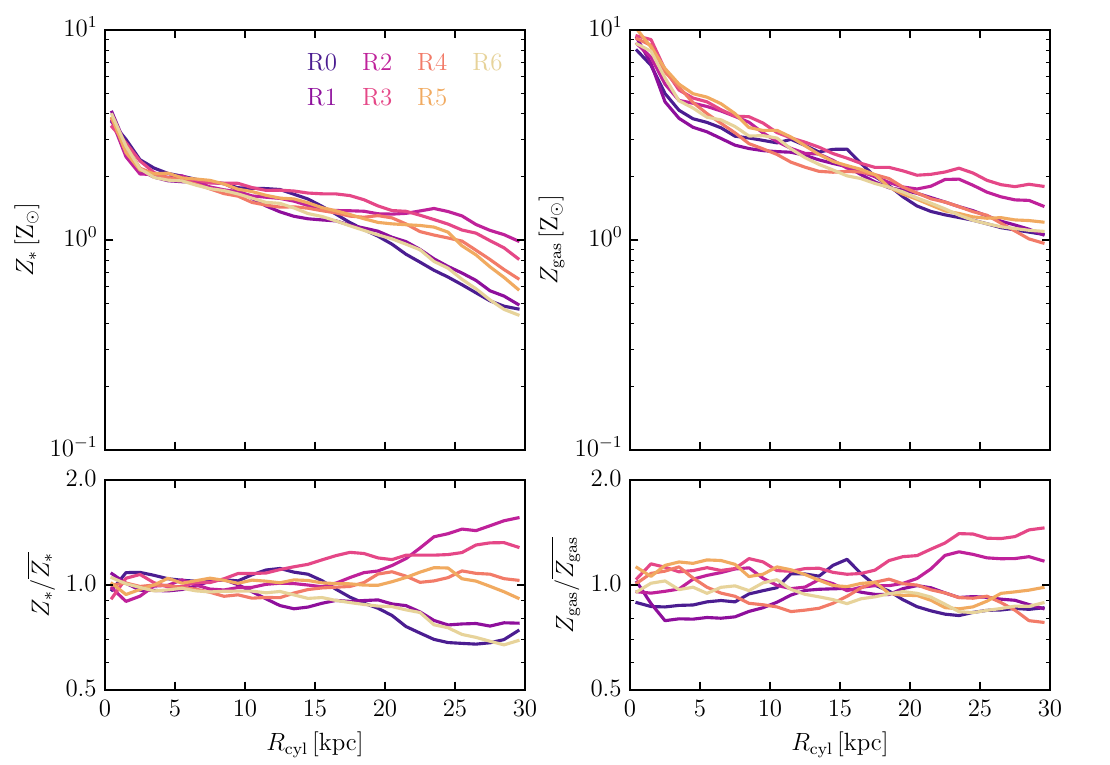}
    \caption{Metallicity profiles of stars (left panels) and gas (right panels) in the disc at $z=0$ in cylindrical coordinates for all seven realisations of Au-6 at the standard resolution L4. In the inner $15\,\mathrm{kpc}$ the stellar metallicity profile is essentially identical, at larger radii it varies from $0.5\,\mathrm{Z_\odot}$ to $\mathrm{Z_\odot}$. The gas metallicity shows slightly larger scatter in the inner parts and slightly less scatter in the outer parts, compared to the stellar metallicity profiles.}
    \label{fig:metallicity_profiles}
\end{figure*}

\section{Global properties of galaxies and their halo}

We show a summary of the global properties of the main galaxy and its halo in all $7$ realisations at $z=0$ in Table~\ref{tab:lvl4}. The total mass of the halo is robust to one per cent. The stellar mass of the central galaxy, and its decomposition into stellar bulge and disc components, change by $5\%$ or less between the realisations. To define the bulge and extended disc we follow \citet{Marinacci2014} and fit a S\'{e}rsic profile for the bulge and an exponential profile for the disc to the stellar surface density profile. Similarly, the size and height of the stellar disc, as well as its total angular momentum, vary only by $\approx 10\%$ and the stellar velocity dispersions at a radius of $8\,\mathrm{kpc}$, as a proxy of the dynamical state of the stellar disc, only vary by less than $5\%$ between realisations. Here, we compute the size and height of the stellar disc by fitting exponentials to the radial and vertical profile of the stellar surface density. We have used this scale radius to compute the stellar mass of the disc.

Notably, the total stellar mass in the halo (at $50\,\mathrm{kpc} < R_\mathrm{3D} < R_\mathrm{200c}$) varies significantly. Here, we include satellite galaxies in the total stellar mass in the halo. The variability arises due to differences in the timing of satellite orbits between realisations which can result in satellites being outside the cut radius of $50\,\mathrm{kpc}$ in some realisations, and inside in others. The other notable variance is in the average star formation rate in the last Gyr, which varies by $\sim30\%$ around the mean value between the realisations.

For a first visual impression of the differences between the galaxies formed in the different realisations we show face-on and edge-on stellar light projections at $z=0$ in Figure~\ref{fig:stellarlight}. We compute these orientations from the angular momentum vector of the stellar disc. The directions of the total angular momentum vectors of the stars in the central galaxy deviate by less than $5^\circ$ from the mean direction. The galaxies look overall very similar. They have a similar size and height, and are all dominated by young, blue stars in the outskirts. They also visually have similarly sized bulges. Interestingly, the internal structure of the stellar disc varies. In particular, R1 features a prominent non-axisymmetric bar-like structure, while R2 looks close to perfectly circular in the centre. The other realisations show mild non-axisymmetric structures somewhere in between R1 and R2. The formation of a bar in only one realisation (R1) indicates that there is intrinsic stochasticity in the dynamical properties of this galactic disc, due to it being marginally bar-unstable (see e.g. \citealt{SellwoodDebattista2009}). In contrast, preliminary tests show that for systems that are highly bar-unstable (for example halo Au-18) bar formation is robust in different random realisations of the same halo (Fragkoudi et al. in prep.). Future detailed analysis of the stochasticity in cosmological simulations will lend further insight into bar formation within the cosmological context.

All galaxies exhibit spiral structures, but they differ in detail. R6 features a clear inner two-armed spiral, the other galaxies show less regular spiral structures. Overall, the global properties of the stellar disc seem robust, but the detailed structural properties vary.

\begin{figure*}
    \centering
    \includegraphics[width=\linewidth]{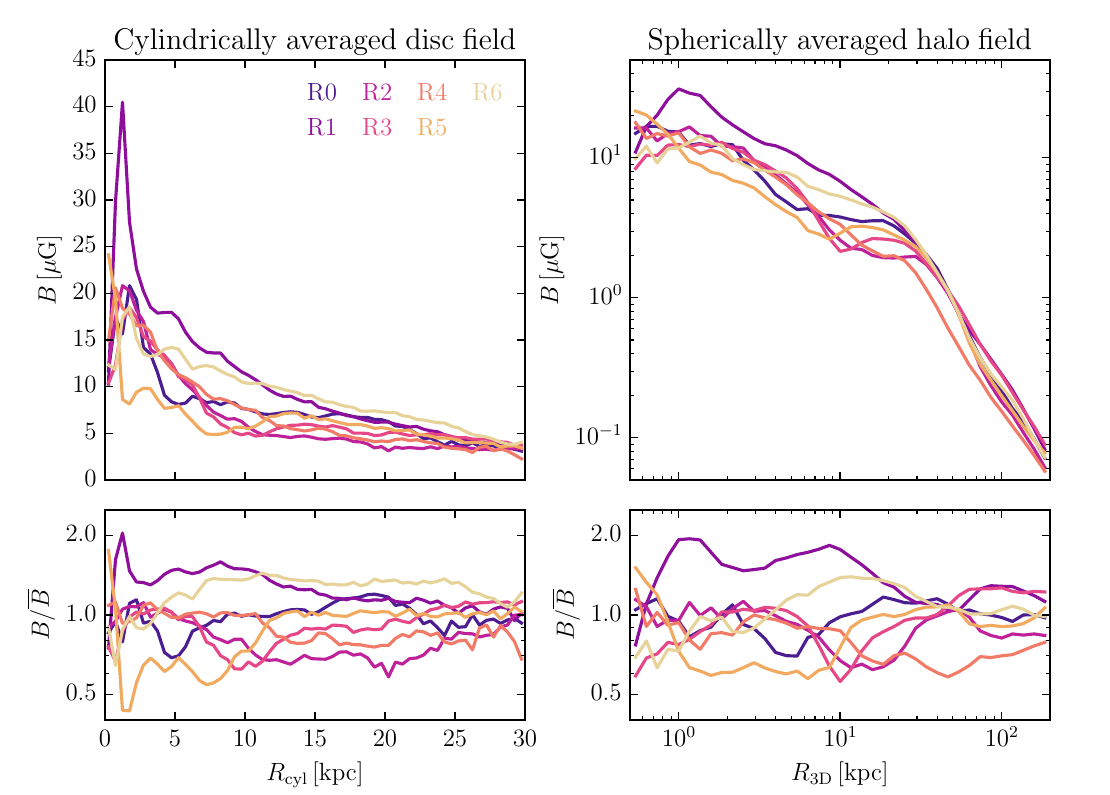}
    \caption{Radial profiles of the strength of the magnetic field at $z=0$ for all seven realisations of Au-6 at resolution L4. The left panel shows the profile in the disc cylindrical coordinates computed from a root-mean-square average of all cells with a height $z<100\,\mathrm{pc}$. The right panel shows the spherically averaged profile out to the virial radius of the halo. The magnetic field strength profiles are consistent within a factor of two, both in the disc as well as in the halo.}
    \label{fig:magnetic_field_strength_profiles}
\end{figure*}

In Figure~\ref{fig:darkmatter} we show projections of the dark matter column density convolved with the local dark matter velocity dispersion rendered using a two-dimensional colormap \citep{Millenium, Pakmor2023MTNG} at $z=0$. The projections are computed along the $z$-axis of the simulation box, so that the orientation of the main halo and the positions of the satellites are directly comparable. The main halo looks very similar in all realisations with similar orientation and elongation. Moreover, the most massive satellite and many of the smaller satellites are present in all halos, at similar positions. We conclude that the dark matter and satellite halos, including their trajectories, are robustly modelled by our simulations. This holds at least until satellites fall deeply into the main halo, similar to simulations that only evolve dark matter and no baryons \citep[see, e.g.][]{Aquarius2}. The deviations in the positions of smaller halos relative to the main halo are caused by errors in the force calculation. Those are directly exposed here, because we add a random shift vector to the global domain at every domain decomposition to decorrelate force errors over time. This reduces the overall force error \citep{Gadget4}, and also leads to different force errors in the different realisations.

As a more quantitative measure of the matter distribution in the halo at $z=0$, we show circular rotation curves of all realisations in Figure~\ref{fig:rotationcurve} (top panel) and their deviation relative to the mean rotation curve of all realisations (bottom panel). The rotation curve directly measures the total enclosed mass at a given radius. Its overall shape is very similar between realisations. The radius where the circular rotation curve reaches its maximum varies only by a few kpc around $15\,\mathrm{kpc}$. The value of the maximum varies only by a few percent, from $195\,\mathrm{km/s}$ to $215\,\mathrm{km/s}$. In summary, the mass distribution in the halo at $z=0$ is very similar for all realisations, and only varies slightly within the central galaxy.

To better quantify the difference in the stellar component of the main galaxy we show the star formation histories of all galaxies in the top panel of Figure~\ref{fig:sfh}, the relative deviation from the mean star formation rate in its middle panel, and the time evolution of the relative deviation of the stellar mass from the mean stellar mass in the bottom panel. Here, we include all star particles bound directly to the main halo at $z=0$ (that is, we exclude star particles bound to satellite galaxies). The star formation histories are essentially identical before $z=2$ and very similar before $z=1$. The initial deviation might be connected to either a significant merger just before $z=2$ or the formation of a disc. After $z=1$ they oscillate around a mean level of $\sim10\,\mathrm{M_\odot/yr}$ until $z\sim0.4$ and then all decline steeply towards $z=0$. At this time, they have dropped by more than a factor of $5$ to a mean star formation rate of only $2\,\mathrm{M_\odot/yr}$, with a significant variance of $1\,\mathrm{M_\odot/yr}$. This variance is driven by differences in the gas accretion rate on the galaxy. Even though the relative difference at $z=0$ is quite large, this has almost no effect on the total stellar mass of the galaxy, because it is dominated by stars formed at earlier times when the relative differences between the different simulations are much smaller. The total stellar mass only deviates on the level of a few per cent at all times.

\begin{figure*}
    \centering
    \includegraphics[width=\linewidth]{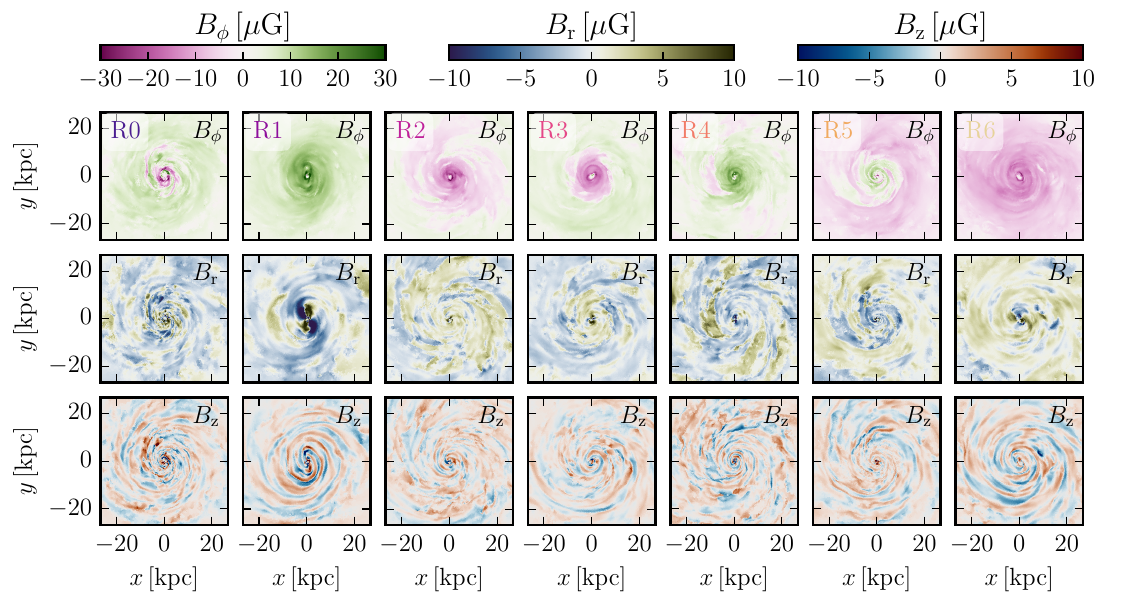}
    \caption{Slices of the different components of the magnetic field in cylindrical coordinates at $z=0$ for all seven realisations of Au-6 at resolution L4. The basic magnetic field properties are remarkably similar for all galaxies. They are dominated by an ordered azimuthal field and show a slightly more chaotic radial field, and a much less ordered vertical field with many field reversals. The detailed structure, in particular the number of field reversals in the azimuthal field, seems stochastic.}
    \label{fig:magnetic_field_slices}
\end{figure*}

\section{Properties of the galactic discs}

Having established that the global properties of the main galaxy and its halo at $z=0$ are very similar for the different realisations, we now focus in more detail on the internal properties and structure of the main galaxy. We first show profiles of the stellar surface density and their relative deviation from the mean profile over all realisations in the left panels of Figure~\ref{fig:stellar_density_profiles}. Here we conservatively include all star particles at a height $\left|z\right| < 30\,\mathrm{kpc}$ in the main halo, though the profiles are completely dominated by the disc. The surface density profiles are all very similar, in normalisation as well as in shape, and for both the bulge in the inner few kpc and the disc at larger radii. 

To look at the full halo, we show cumulative stellar mass profiles of all realisations in the right panels of Figure~\ref{fig:stellar_density_profiles}. Consistent with Table~\ref{tab:lvl4} we see that the total stellar mass in the halo is very similar. Notably, in R1 the stellar mass is more concentrated in the very centre of the galaxy. This central concentration is likely directly connected to the bar \citep{Fragkoudi2024} that is prominently featured only in this realisation (see Figure~\ref{fig:stellarlight}).

We show stellar metallicity profiles and their relative deviation from the mean profile in the left panels of Figure~\ref{fig:metallicity_profiles}. In the inner $15\,\mathrm{kpc}$ the stellar metallicity profile only deviates by a few per cent. At larger radii the deviation is significantly larger. In the very outskirts of the stellar disc at $30\,\mathrm{kpc}$ the stellar metallicity varies from $0.5\,\mathrm{Z_\odot}$ to $\mathrm{Z_\odot}$. The deviation from the mean stellar metallicity profile (see lower left panel of Figure~\ref{fig:metallicity_profiles}) in the outskirts strongly correlates with the deviation from the mean stellar surface density profile (see lower left panel of Figure~\ref{fig:stellar_density_profiles}). Higher stellar surface density leads to higher stellar metallicity and vice versa.

Having shown that the stellar discs are similar for the different realisations, we next look at gas metallicity profiles at $z=0$ in the right panels of Figure~\ref{fig:metallicity_profiles}. The gas metallicity profiles have slightly larger scatter in the inner parts of the disc compared to the stellar metallicity profiles, and slightly smaller scatter in the outer parts. The scatter is still below $20\%$ at all radii.

We then look at the magnetic fields in the mid-plane of the disc at $z=0$ as a proxy to compare how similar the gas discs are. Magnetic fields affect the gas dynamics and are an important tracer of the state of the gas in both the disc \citep{Pakmor2014,Pakmor2017,Pakmor2018,Whittingham2021,Whittingham2023,Pakmor2024} and the circum-galactic medium \citep{Pakmor2020,vandeVoort2021}. Note that this comparison is only really meaningful because the stellar discs, which dominate the potential in the galaxy and therefore the gas dynamics in the galaxy as well, are so similar. 

We show the magnetic field strength profile in the disc in Figure~\ref{fig:magnetic_field_strength_profiles} (left panels) and in the full halo (right panels) at $z=0$. The magnetic field strength profiles in the discs are similar. At a given radius the magnetic field strength deviates by less than a factor of $2$ around the mean profile. R1 is again an interesting outlier, that has a significantly stronger magnetic field in the very centre of galaxy. This is likely a result of the higher stellar mass and therefore deeper potential in the very centre of the galaxy and possibly also connected to the presence of a bar as discussed above.

The magnetic field strength in the halo outside the galaxy, that is, at radii larger than $R_\mathrm{3D}\gtrsim 30\,\mathrm{kpc}$, also agrees well for the different realisations. Most of the realisations deviate less than $20\%$ from the mean magnetic field strength profile. Notably, R4 has a slightly lower magnetic field strength in the halo, but approaches the mean profile at the edge of the halo.

In the Auriga galaxies magnetic fields are quickly amplified when the galaxy first forms at high redshift. This turbulent dynamo saturates before a stable gas disc forms around $z=2$ \citep{Pakmor2017,Pakmor2024}. The rotation in the gas disc then orders and further amplifies the magnetic field. At $z=0$ it is typically dominated by a large-scale ordered field \citep{Pakmor2018,Pakmor2024}. Therefore, the structure of the magnetic field at $z=0$ can also provide important diagnostics for the evolution of the galaxy and its gas disc. 

We show slices of the magnetic field components in cylindrical coordinates in the disc at $z=0$ for all realisations in Figure~\ref{fig:magnetic_field_slices}. As a first impression, we see that all three components (azimuthal magnetic field in the top row, radial magnetic field in the middle row, and vertical magnetic field in the bottom row) show qualitatively similar structures in all realisations. All galaxies feature large scale ordered azimuthal magnetic fields that are about $3$ times stronger than the radial and vertical magnetic fields. 

Field reversals in the azimuthal field are rare, the galaxies only have either zero or one field reversals. The sign of the ordered field is not a robust outcome, however. This is not surprising, as the equations of ideal MHD are agnostic about the sign of the magnetic field. Therefore, either ordered configuration is physically indistinguishable.

The vertical magnetic fields (shown in the bottom row of Figure~\ref{fig:magnetic_field_slices}), in contrast, are not ordered on large scales. They all show signs of being wound up from an initial chaotic small-scale field. This leads to a field that is ordered along the azimuth with many field reversals. Its strength is also similar for all galaxies. 

The radial magnetic field (middle row of Figure~\ref{fig:magnetic_field_slices}) has an intermediate structure. It shows some hints of large scale ordering, but still has most of the small-scale structure and many field reversals of the vertical field. Its strength is also again similar for all realisations. One notable outlier is again R1, which has a significantly stronger radial field in the centre of the galaxy, and shows the dipolar pattern typically associated with bars.

\section{Satellite galaxies}

In addition to the central galaxy, a cosmological zoom-in simulation also contains a number of smaller galaxies, some of them satellites of the main galaxy. The latter are interesting in and of themselves and as a diagnostic for the galaxy evolution model \citep{Simpson2018,Grand2021}. They also interact with their central host galaxy. They shape its stellar halo and create stellar streams \citep{Monachesi2016,Monachesi2019,Riley2024,Shipp2024,Vera2025}, they also influence the disc of the central galaxy, and they contribute to substructures in the disc and inner halo \citep{Gomez2016, Gomez2016b, Simpson2019, Gargiulo2019}. Therefore, it is crucial to understand how similar the satellites are between realisations when they fall into the halo, to determine how robust their evolution is until their possible disruption.

\begin{figure*}
    \centering
    \includegraphics[width=\linewidth]{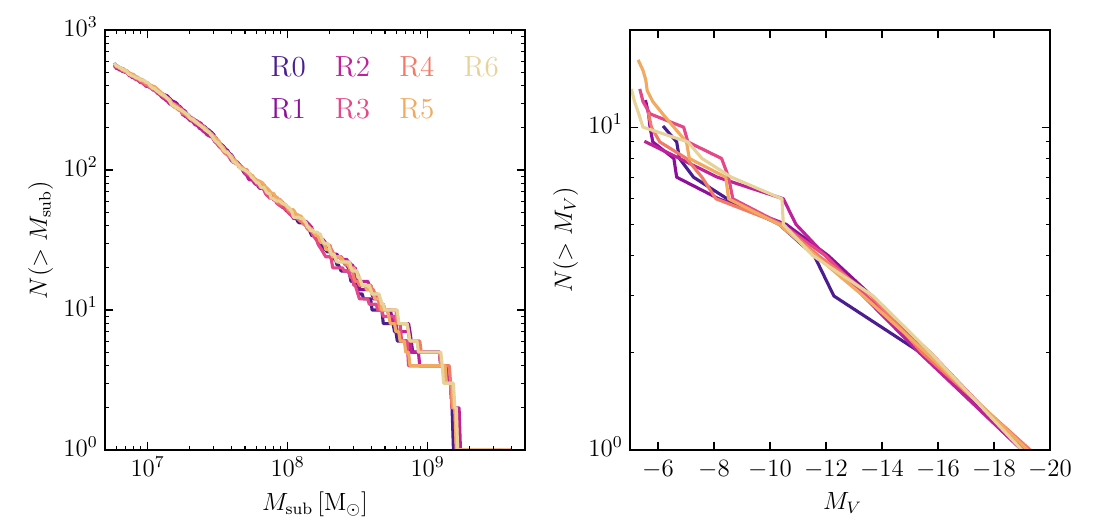}
    \caption{Satellite total mass function (left panel) and V-band luminosity function (right panel)  at $z=0$ for all seven realisations of Au-6 at resolution L4. The mass functions are essentially identical. The luminosity function only shows significant scatter for $M_\mathrm{V} > -10$, with a single outlier at $M_\mathrm{V}=-12$ for R0.}
    \label{fig:satellites}
\end{figure*}

Figure~\ref{fig:satellites} shows the satellite mass function (left panel), that shows the total mass of all satellites at $z=0$, and the satellite luminosity function (right panel), that shows the $V$-band luminosity of all satellites at $z=0$. As discussed in detail in \citet{Grand2021}, both show very good agreement between the different realisations and we conclude that they are robust outcomes of our simulations.

\begin{figure*}
    \centering
    \includegraphics[width=\linewidth]{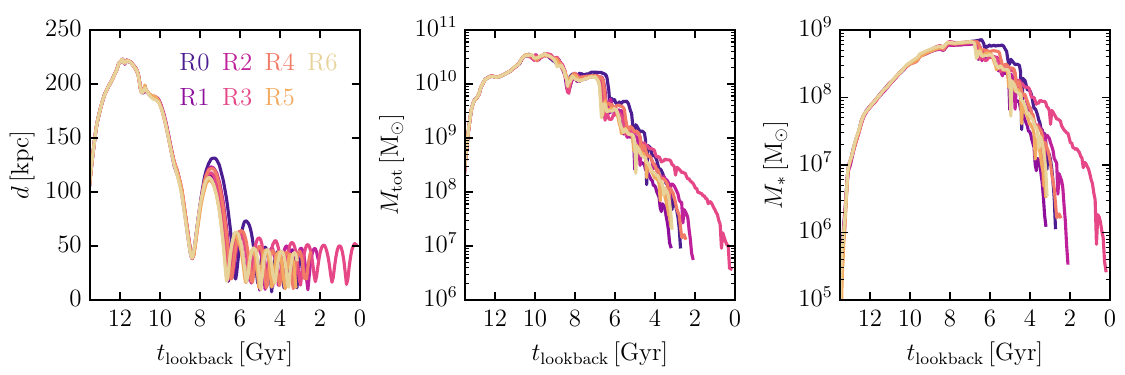}
    \caption{Trajectories of one relatively massive satellite that is destroyed before $z=0$ for all seven realisations of Au-6 at resolution L4. The panels show the time evolution of the distance to the centre of the galaxy (left panel), total bound mass (middle panel), and total bound stellar mass (right panel). The satellite falls in with essentially identical properties. The longer it orbits the main galaxy, the more its properties deviate between the realisations.}
    \label{fig:satellite_trajectory}
\end{figure*}

In addition to looking at satellite galaxies that survive until $z=0$ it is interesting to compare satellites that are destroyed. As an example, we show in Figure~\ref{fig:satellite_trajectory} the evolution of one satellite that is eventually destroyed. As a representative case, we select the third most massive (by stellar mass) satellite among those disrupted before $z=0$ in R0. We then match it to the other realisations by tracing back the $100$ most bound dark matter particles of each satellite at the time it reaches its peak mass to the initial conditions \citep[see also,][]{Riley2024}. We then compute the centre of mass of these particles, and find the satellite that most closely matches the centre of our reference satellite selected from R0.

Figure~\ref{fig:satellite_trajectory} shows the time evolution of the distance of this satellite to the centre of the main galaxy (left panel), the time evolution of the total mass of the satellite (middle panel), and the time evolution of the stellar mass of the satellite (right panel). The initial trajectory of the satellite when it falls into the main halo until it reaches its first closest encounter (peri-centre) and its total mass and stellar mass at the time of infall are essentially identical. The trajectories of the satellite in the different realisations start to deviate slightly after their second apo-centre passage, and spread more and more with every further orbit. The periods of the orbit of the satellite remain similar in the different realisations, but they deviate enough to quickly change the phase of the orbit. The realisations of the same halo will potentially allow us to understanding which properties of the parent halo are the main source of the differences in the satellite trajectories.

The time when the satellite is eventually destroyed, that is when it is not found by subfind anymore as bound object, varies from $t_\mathrm{lookback}=3\mathrm{Gyr}$ to $t_\mathrm{lookback}=2\mathrm{Gyr}$ between the different realisations, with one notable outlier (R3). In this realisation the satellite survives more than $2\,\mathrm{Gyr}$ longer than in the realisation where it survives for the second longest time.
The total mass and stellar mass of the satellite both initially evolve very similarly between the different realisations, which means that mass stripping is robust for a fixed numerical resolution. After several orbits, at $t_\mathrm{lookback}\approx 5\,\mathrm{Gyr}$ the realisation (R3) in which the satellite survives significantly longer starts to diverge, as the satellite is stripped less quickly than in the other realisations. This is possibly a consequence of a slightly larger first apocentre distance, but more work will be required to understand this difference in detail. The other realisations evolve very similarly all the way to the destruction of the satellite.

\begin{figure*}
    \centering
    \includegraphics[width=\linewidth]{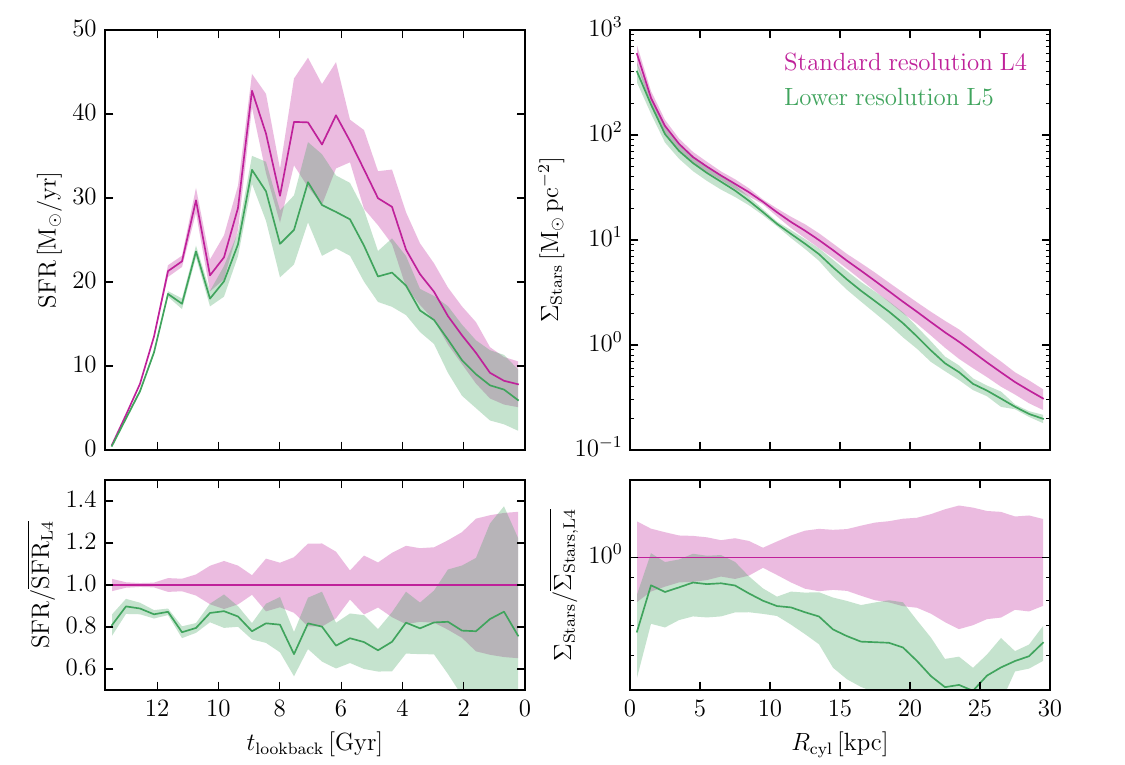}
    \caption{Star formation histories (top left panel) and stellar mass surface density profiles (top right panel) of the main galaxy for all seven realisations of Au-6 at standard resolution L4 (purple lines) and $8$ times lower resolution L5 (green lines). The shaded regions show interpolated $16\%$ and $84\%$ percentile bands around the mean profile (solid lines). The bottom panels show the same relative to the mean of all simulations at L4. The differences between the two resolution levels in all quantities are significantly larger than the variations between realisations at fixed resolution.}
    \label{fig:resolution_comparison}
\end{figure*}

\begin{figure}
    \centering
    \includegraphics[width=\linewidth]{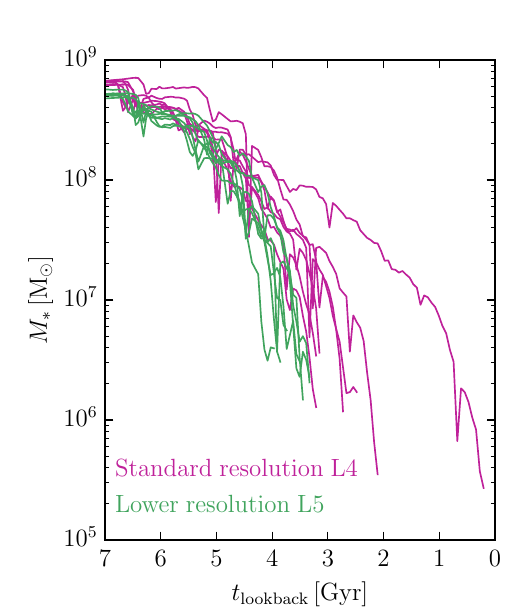}
    \caption{Time evolution of the stellar mass of one matched satellite galaxy of the main galaxy for all seven realisations of Au-6 at the standard resolution L4 (purple lines, the same as in the right panel of Figure~\ref{fig:satellite_trajectory}) and $8$ times lower resolution L5 (green lines). Similar to the main galaxy the peak stellar mass of the satellite galaxy is lower for the lower resolution simulation. The satellite is destroyed slightly earlier on average in the lower resolution simulation.}
    \label{fig:resolution_comparison_satellite}
\end{figure}

\begin{table*}
\textbf{Realisations at lower resolution (L5)}
\begin{center}
\setlength\tabcolsep{2pt}
\begin{tabular}{ c c c c c c c c c c c c c }
\hline
Realisation & $M_{200\mathrm{c}}$ & $M_*$ & $M_{*,R>50\mathrm{kpc}}$ & $\dot{M}_{1\mathrm{Gyr}}$ & $V_\mathrm{c,max}$ & $M_\mathrm{disc}$ & $M_\mathrm{bulge}$ & $R_\mathrm{disc}$ & $H_\mathrm{disc}$ & $\sigma_\mathrm{r}$ & $\sigma_\mathrm{z}$ & $\left|L_*\right|$ \\
& $[10^{12}\rmn{M}_\odot]$ & $[10^{10}\rmn{M}_\odot]$ & $[10^{10}\rmn{M}_\odot]$ & $[\rmn{M}_\odot / \rmn{yr}]$ & $[\rmn{km/s}]$ & $[10^{10}\rmn{M}_\odot]$ & $[10^{10}\rmn{M}_\odot]$ & $[\rmn{kpc}]$ & $[\rmn{kpc}]$ & $[\rmn{km/s}]$ & $[\rmn{km/s}]$ & $[\rmn{10^{18}\rmn{M}_\odot\,kpc\,km/s}]$ \\
\hline
0 & $1.04$ & $4.37$ & $0.22$ & $1.5$ & $204$ & $0.35$ & $3.78$ & $3.7$ & $1.10$ & $46.8$ & $39.0$ & $4.4$\\
1 & $1.04$ & $4.07$ & $0.29$ & $0.3$ & $198$ & $0.35$ & $3.40$ & $4.0$ & $1.04$ & $43.5$ & $37.1$ & $4.0$\\
2 & $1.03$ & $4.17$ & $0.31$ & $2.9$ & $204$ & $0.35$ & $3.49$ & $3.7$ & $1.14$ & $50.0$ & $40.1$ & $4.0$\\
3 & $1.03$ & $4.74$ & $0.60$ & $3.2$ & $203$ & $0.35$ & $3.74$ & $4.0$ & $1.00$ & $46.5$ & $38.5$ & $4.7$\\
4 & $1.05$ & $4.47$ & $0.27$ & $1.7$ & $204$ & $0.36$ & $3.76$ & $3.6$ & $1.04$ & $45.6$ & $38.7$ & $4.3$\\
5 & $1.02$ & $4.16$ & $0.24$ & $3.3$ & $200$ & $0.37$ & $3.57$ & $4.4$ & $1.26$ & $47.5$ & $41.7$ & $4.6$\\
6 & $1.03$ & $3.77$ & $0.29$ & $1.0$ & $195$ & $0.37$ & $3.13$ & $4.9$ & $1.41$ & $44.1$ & $37.8$ & $4.1$\\
\hline
mean & $1.03$ & $4.25$ & $0.32$ & $2.0$ & $201$ & $0.36$ & $3.55$ & $4.0$ & $1.14$ & $46.3$ & $39.0$ & $4.3$\\
$\sigma$ & $0.01$ & $0.31$ & $0.13$ & $1.2$ & $3.7$ & $0.01$ & $0.24$ & $0.5$ & $0.15$ & $2.2$ & $1.5$ & $0.3$\\
$\sigma$ / mean $[\%]$ & $0.9$ & $7.3$ & $40.6$ & $58.3$ & $1.8$ & $2.7$ & $6.7$ & $11.4$ & $12.8$ & $4.7$ & $3.9$ & $6.6$\\
\hline
$\cfrac{ \rmn{mean} - \rmn{mean_{L4}} }{ \rmn{mean_{L4}} } [\%]$ & $-0.8$ & $-21.1$ & $-39.4$ & $-17.5$ & $-2.7$ & $-14.8$ & $-18.8$ & $-9.5$ & $28.6$ & $-2.6$ & $-1.1$ & $-33.6$\\
\hline
\end{tabular}
\end{center}
\caption{Global properties of our galaxy in all $7$ realisations at $z=0$ for the lower resolution simulations (L5). The columns show from left to right the mass of the halo $M_{200\mathrm{c}}$, the stellar mass of the main subhalo $M_*$, the stellar mass in the halo at a distances larger than $50\,\mathrm{kpc}$ $M_{*,R>50\mathrm{kpc}}$, the star formation rate in the last Gyr $\dot{M}_{1\mathrm{Gyr}}$, the maximum circular velocity $V_\mathrm{c,max}$, the mass of the stellar disc $M_\mathrm{disc}$ and bulge $M_\mathrm{bulge}$, the scale radius $R_\mathrm{disc}$ and height $H_\mathrm{disc}$ of the stellar disc, the radial $\sigma_\mathrm{r}$ and vertical $\sigma_\mathrm{z}$ velocity dispersion in the stellar disc at a radius of $8\,\mathrm{kpc}$, and the total angular momentum of the stellar disc $\left|L_*\right|$. The last three rows below show the mean value, variance, and the variance relative to the mean for all quantities over all $7$ realisations. The lowest row shows the relative difference of the mean to the mean of the simulations at the standard resolution (L4) shown in Table~\ref{tab:lvl4}. Resolution in particular changes stellar masses significantly more than the intrinsic variation.}
\label{tab:lvl5}
\end{table*}

\section{Intrinsic variance and numerical resolution}

Having quantified the intrinsic variance of the Auriga model at fixed numerical resolution, an obvious next step is to extend this comparison to simulations with different numerical resolution. In Figure~\ref{fig:resolution_comparison} we compare two sets of simulations with $7$ realisations each. We use the realisations discussed in this paper at the standard resolution of Auriga (L4), and a set of simulations with $8$x worse mass resolution (L5) for all mass components, that is for gas, stars, and dark matter. We compare star formation histories (left panels of Figure~\ref{fig:resolution_comparison}) and profiles of the stellar mass surface density at $z=0$ (right panels of Figure~\ref{fig:resolution_comparison}). The bands show interpolated $16\%$ and $84\%$ percentiles. We immediately see that for the first $10\,\mathrm{Gyr}$ of cosmic evolution the star formation rate is systematically lower by $\approx20\%$ in the simulations with worse mass resolution than in the standard resolution simulations. The difference is much larger than the intrinsic variability at fixed resolution. This difference is consistent with previous results for the Auriga model \citep{Auriga, Grand2021} and similarly the IllustrisTNG model \citep{Pillepich2018}. In the last few Gyr, when the star formation rate drops significantly in all simulations and the relative fractional difference between simulations at the same resolution increases (see also Figure~\ref{fig:sfh}), the distributions overlap. However, the mean is still shifted systematically to lower star formation rates at lower resolution.

Moreover, because the star formation rate is lower over the whole evolution of the galaxies in the lower resolution simulations, their stellar mass at $z=0$ is also systematically lower. Looking at the stellar surface density profiles at $z=0$ (right panels of Figure~\ref{fig:resolution_comparison}) we can see that the deficit of stellar mass in the lower resolution simulations is distributed over the full galaxy and not limited to either the central bulge or the extended disc. The systematically different stellar surface density profiles also imply that the dynamical state of the stellar disc is different, which modifies the conditions for internal processes of the disc, for example for the formation of a bar \citep{Fragkoudi2021, Fragkoudi2024}. As a result, it becomes difficult to disentangle genuine resolution effects on specific physical processes, such as bar formation, from the broader structural differences introduced by resolution. This limits the interpretability of resolution studies not only when they focus on the disc but also for all other properties of the satellites and halo that might be affected by the central disc galaxy and its feedback processes.

We see a similar effect for the satellite galaxy in Figure~\ref{fig:resolution_comparison_satellite} (the same that is shown in Figure~\ref{fig:satellite_trajectory}), where we show the time evolution of the stellar mass of a satellite matched between all realisations of all simulations of both resolution levels. We see that the satellite galaxy has a systematically lower stellar mass at all times before it falls into the main galaxy in the lower resolution simulations (L5). This systematic difference essentially disappears when the satellite galaxy is starting to get stripped and loses stellar mass. The satellite tends to be destroyed marginally faster in the lower resolution simulation. For most other satellites (not shown), however, their destruction time is not significantly different for simulations at different resolution. The different resolution dependence of the destruction time might depend on details of the orbit of the satellite and the mass and other properties of the stellar disc of the central galaxy that change as well. Even if the satellite galaxy gets destroyed at the same time in different resolution simulations, it will still have contributed less stellar mass to the halo in the lower resolution simulations, biasing the stellar halo systematically in the lower resolution simulation in a similar way as the central galaxy. The maximum stellar mass of the satellites in the lower resolution (L5) runs is lower by $20\%$ compared to the standard resolution (L4) runs. This ratio is the same as ratio of the stellar masses of the central galaxies at $z=0$ for the different resolutions.

Finally, we summarise the same global properties of the main galaxy and its halo at $z=0$ shown in Table~\ref{tab:lvl4} for the standard (L4) resolution realisations also for the lower resolution (L5) realisations in Table~\ref{tab:lvl5}. Comparing the mean and variance of the stellar properties of the $z=0$ stellar disc of the standard resolution realisations (L4) and lower resolution realisations (L5), we see that for most stellar properties (stellar mass, disc mass, bulge mass, disc height) the differences introduced by changing the resolution is significantly larger than the intrinsic variance of the model at fixed resolution. Notably, there are hints that the intrinsic variance decreases with higher resolution. This is probably a result of the reduced impact of each individual random number in the higher resolution simulations. The latter has more star-forming cells in the same galaxy. It draws an independent random number for each of those cells to determine if it creates a star particle or wind particle, and each star particle and wind particle is less massive. Therefore, the importance of each random number is lower in the higher resolution simulation, and extreme outliers become rarer.

The IllustrisTNG model also shows a reduction of variability at better numerical resolution when comparing pairs of Milky Way-like galaxies in different realisations \citep{Genel2019}. However, this reduction of variability mostly disappears for lower mass galaxies. Moreover, \citet{Genel2019} demonstrate that for sufficiently high numerical resolution the variability in the IllustrisTNG model for these low-mass galaxies is not primarily driven by random numbers in the subgrid model.

\section{Summary and discussion}

We have quantified the intrinsic variability of the Auriga model as applied to a cosmological zoom-in simulation of a Milky Way-like galaxy. We showed in Figures~\ref{fig:darkmatter} and \ref{fig:rotationcurve} that the dark matter structures, including the shape and total mass profile of the main halo and the positions of the satellites at $z=0$ are robust, with relative deviations from the mean only on the percent level.

Moreover, we showed that the global properties of the stellar disc of the main galaxy at $z=0$, including its total mass and star formation history (see Figure~\ref{fig:sfh}) and its stellar surface density profile (see Figure~\ref{fig:stellar_density_profiles}) are robust with relative deviations from the mean profile smaller than $\approx 30\%$. Variations in global quantities are even smaller, that is the total stellar mass of the central galaxy, of its bulge and extended disc, as well as the stellar mass of satellite galaxies deviate by only $\approx 5\%$. Similarly, the strength (see Figure~\ref{fig:magnetic_field_strength_profiles}) and structure (see Figure~\ref{fig:magnetic_field_slices}) of the magnetic field in the disc and circum-galactic medium, are a robust outcome of the simulation and deviate by less than a factor of two locally. Therefore they are useful properties for comparison to observations. We also showed that the satellite population of the main halo at $z=0$ (see Figure~\ref{fig:satellites}) and the properties and evolution of individual satellites and their stripping and potential eventual destruction are robust in the Auriga model, at least until first pericentre passage (see Figure~\ref{fig:satellite_trajectory}).

All properties of the central galaxy of all realisations are much more similar than the initial set of $30$ galaxies of the Auriga project \citep{Auriga} that are all in relatively isolated Milky Way-mass halos. Thus, their formation histories seem to robustly determine their global properties, much more than the intrinsic variability of the Auriga model and the numerical schemes. This strengthens comparisons between different halos of the Auriga project with the aim to connect physical properties of galaxies to their formation history. It will also allow us to better understand how different physical properties of galaxies are correlated and caused by each other in our set of realisations that all have the same formation history.

Finally, we showed in Figure~\ref{fig:resolution_comparison} and Figure~\ref{fig:resolution_comparison_satellite} that changing the numerical resolution of the simulation by a factor of $8$ leads to significant systematic changes, which are much bigger than the intrinsic variability of the Auriga model at fixed resolution. The small intrinsic variance of the Auriga model makes it attractive to study physical model variations at fixed resolution \citep[see, e.g.][]{Pakmor2017,Buck2019,Montero2024,vandeVoort2021}. However, the systematic changes to the galaxy properties with changing numerical resolution not only limits proper resolution studies, but also limits going to higher resolution without recalibrating a model.

Recent studies looking into the intrinsic variability of galaxy formation models have shown that it is important to understand the robustness of a model to interpret its results, and that the robustness is significantly different for different models \citep{Keller2019, Genel2019, Davies2021, Borrow2023}. A detailed quantitative comparison between the intrinsic variability of different models is hard because most published work limits quantitative discussions to the total stellar mass.

The four different feedback models discussed in \citet{Keller2019} show significant differences. They focus on the evolution of stellar mass and show that their "Superbubble" feedback has the largest variance in stellar mass of $10\%$ at $z=2$ over $128$ realisations and of $20\%$ between a pair of simulations evolved further to $z=0.8$. For this model they also find differences in the stellar surface density at $z=0.8$ of almost an order of magnitude. This level of variance is significantly larger than for the Auriga model. Their other feedback models have comparable variance at $z=2$ and lower variance at $z=0.8$, but at the cost of inefficient feedback that is essentially consistent with no feedback.

The comparison of a set of large volume simulations with the IllustrisTNG model \citep{Genel2019} showed that massive galaxies can end up with completely different morphologies. This is a result of the very stochastic radio-mode AGN feedback model of IllustrisTNG. The Auriga model, which is very similar to the IllustrisTNG model in terms of stellar feedback, features a radio-mode AGN feedback model that is much more gentle and less stochastic \citep{Auriga}. The IllustrisTNG simulations also show roughly $5\%$ variance in the value of the maximum of the circular velocity curve, averaged over about $100$ Milky Way-like galaxies. This is roughly consistent with our results for the Auriga model. The stellar mass of matched Milky Way-like galaxies between the IllustrisTNG boxes varies by $10\%$ at $z=0$ for the highest resolution box with roughly comparable mass resolution to the standard Auriga (L4) resolution, slightly larger than the variance ($\lesssim 5\%$) we find for our single galaxy.

Similar studies with the EAGLE model \citep{Davies2021} and the SWIFT-EAGLE model \citep{Borrow2023} show variance on the $1\%$ level in the halo mass at $z=0$ for Milky Way-like galaxies, similar to our simulations. Moreover, they show an average variance in the stellar mass of Milky Way-like galaxies on the $10\%$ level (EAGLE) and between $5\%$ and $10\%$ (SWIFT-EAGLE), again slightly larger than for our single galaxy.

These differences should motivate us to include similar robustness and convergence studies when designing new galaxy formation models, in particular for cosmological zoom simulations. Ideally this would allow us to increase the robustness of a model not only at fixed resolution, but also when changing the numerical resolution. As one specific consequence of the study presented here, we will introduce a new approach that improves the stellar mass resolution in cosmological galaxy simulations, but avoids systematically changing galaxy properties at the same time \citep{Pakmor2025b}.

\section*{Acknowledgements}

RB is supported by the UZH Postdoc Grant, grant no. FK-23116 and the SNSF through the Ambizione Grant PZ00P2\_223532.
RJJG acknowledges support from an STFC Ernest Rutherford Fellowship (ST/W003643/1).
FvdV is supported by a Royal Society University Research Fellowship (URF\textbackslash R1\textbackslash191703 and URF\textbackslash R\textbackslash241005). FF is supported by a UKRI Future Leaders Fellowship (grant no. MR/X033740/1).
The authors gratefully acknowledge the Gauss Centre for Supercomputing e.V. (www.gauss-centre.eu) for compute time on the GCS Supercomputer
SUPERMUC-NG at Leibniz Supercomputing Centre (www.lrz.de).

\section*{Data Availability}
The data underlying this article will be shared on reasonable request to the corresponding author.



\bibliographystyle{mnras}

\begin{thebibliography}{}
\makeatletter
\relax
\def\mn@urlcharsother{\let\do\@makeother \do\$\do\&\do\#\do\^\do\_\do\%\do\~}
\def\mn@doi{\begingroup\mn@urlcharsother \@ifnextchar [ {\mn@doi@}
  {\mn@doi@[]}}
\def\mn@doi@[#1]#2{\def\@tempa{#1}\ifx\@tempa\@empty \href
  {http://dx.doi.org/#2} {doi:#2}\else \href {http://dx.doi.org/#2} {#1}\fi
  \endgroup}
\def\mn@eprint#1#2{\mn@eprint@#1:#2::\@nil}
\def\mn@eprint@arXiv#1{\href {http://arxiv.org/abs/#1} {{\tt arXiv:#1}}}
\def\mn@eprint@dblp#1{\href {http://dblp.uni-trier.de/rec/bibtex/#1.xml}
  {dblp:#1}}
\def\mn@eprint@#1:#2:#3:#4\@nil{\def\@tempa {#1}\def\@tempb {#2}\def\@tempc
  {#3}\ifx \@tempc \@empty \let \@tempc \@tempb \let \@tempb \@tempa \fi \ifx
  \@tempb \@empty \def\@tempb {arXiv}\fi \@ifundefined
  {mn@eprint@\@tempb}{\@tempb:\@tempc}{\expandafter \expandafter \csname
  mn@eprint@\@tempb\endcsname \expandafter{\@tempc}}}

\bibitem[\protect\citeauthoryear{{Bieri}, {Pakmor}, {van de Voort}, {Talbot},
  {Werhahn}, {Pfrommer}  \& {Springel}}{{Bieri} et~al.}{2025}]{Bieri2025}
{Bieri} R.,  {Pakmor} R.,  {van de Voort} F.,  {Talbot} R.~Y.,  {Werhahn} M.,
  {Pfrommer} C.,   {Springel} V.,  2025, \mn@doi [arXiv e-prints]
  {10.48550/arXiv.2509.07124}, \href
  {https://ui.adsabs.harvard.edu/abs/2025arXiv250907124P} {p. arXiv:2509.07124}

\bibitem[\protect\citeauthoryear{{Borrow}, {Schaller}, {Bah{\'e}}, {Schaye},
  {Ludlow}, {Ploeckinger}, {Nobels}  \& {Altamura}}{{Borrow}
  et~al.}{2023}]{Borrow2023}
{Borrow} J.,  {Schaller} M.,  {Bah{\'e}} Y.~M.,  {Schaye} J.,  {Ludlow} A.~D.,
  {Ploeckinger} S.,  {Nobels} F. S.~J.,   {Altamura} E.,  2023, \mn@doi
  [\mnras] {10.1093/mnras/stad2928}, \href
  {https://ui.adsabs.harvard.edu/abs/2023MNRAS.526.2441B} {526, 2441}

\bibitem[\protect\citeauthoryear{{Buck}, {Pfrommer}, {Pakmor}, {Grand}  \&
  {Springel}}{{Buck} et~al.}{2020}]{Buck2019}
{Buck} T.,  {Pfrommer} C.,  {Pakmor} R.,  {Grand} R. J.~J.,   {Springel} V.,
  2020, \mn@doi [\mnras] {10.1093/mnras/staa1960}, \href
  {https://ui.adsabs.harvard.edu/abs/2020MNRAS.497.1712B} {497, 1712}

\bibitem[\protect\citeauthoryear{{Crain} \& {van de Voort}}{{Crain} \& {van de
  Voort}}{2023}]{Crain2023}
{Crain} R.~A.,  {van de Voort} F.,  2023, \mn@doi [\araa]
  {10.1146/annurev-astro-041923-043618}, \href
  {https://ui.adsabs.harvard.edu/abs/2023ARA&A..61..473C} {61, 473}

\bibitem[\protect\citeauthoryear{{Dav{\'e}}, {Angl{\'e}s-Alc{\'a}zar},
  {Narayanan}, {Li}, {Rafieferantsoa}  \& {Appleby}}{{Dav{\'e}}
  et~al.}{2019}]{Dave2019Simba}
{Dav{\'e}} R.,  {Angl{\'e}s-Alc{\'a}zar} D.,  {Narayanan} D.,  {Li} Q.,
  {Rafieferantsoa} M.~H.,   {Appleby} S.,  2019, \mn@doi [\mnras]
  {10.1093/mnras/stz937}, \href
  {https://ui.adsabs.harvard.edu/abs/2019MNRAS.486.2827D} {486, 2827}

\bibitem[\protect\citeauthoryear{{Davies}, {Crain}  \& {Pontzen}}{{Davies}
  et~al.}{2021}]{Davies2021}
{Davies} J.~J.,  {Crain} R.~A.,   {Pontzen} A.,  2021, \mn@doi [\mnras]
  {10.1093/mnras/staa3643}, \href
  {https://ui.adsabs.harvard.edu/abs/2021MNRAS.501..236D} {501, 236}

\bibitem[\protect\citeauthoryear{{Davies}, {Pontzen}  \& {Crain}}{{Davies}
  et~al.}{2022}]{Davies2022}
{Davies} J.~J.,  {Pontzen} A.,   {Crain} R.~A.,  2022, \mn@doi [\mnras]
  {10.1093/mnras/stac1742}, \href
  {https://ui.adsabs.harvard.edu/abs/2022MNRAS.515.1430D} {515, 1430}

\bibitem[\protect\citeauthoryear{{Dubois}, {Peirani}, {Pichon}, {Devriendt},
  {Gavazzi}, {Welker}  \& {Volonteri}}{{Dubois}
  et~al.}{2016}]{Dubois2016HorizonAGN}
{Dubois} Y.,  {Peirani} S.,  {Pichon} C.,  {Devriendt} J.,  {Gavazzi} R.,
  {Welker} C.,   {Volonteri} M.,  2016, \mn@doi [\mnras]
  {10.1093/mnras/stw2265}, \href
  {https://ui.adsabs.harvard.edu/abs/2016MNRAS.463.3948D} {463, 3948}

\bibitem[\protect\citeauthoryear{{Fragkoudi}, {Grand}, {Pakmor}, {Springel},
  {White}, {Marinacci}, {Gomez}  \& {Navarro}}{{Fragkoudi}
  et~al.}{2021}]{Fragkoudi2021}
{Fragkoudi} F.,  {Grand} R.~J.~J.,  {Pakmor} R.,  {Springel} V.,  {White}
  S.~D.~M.,  {Marinacci} F.,  {Gomez} F.~A.,   {Navarro} J.~F.,  2021, \mn@doi
  [\aap] {10.1051/0004-6361/202140320}, \href
  {https://ui.adsabs.harvard.edu/abs/2021A&A...650L..16F} {650, L16}

\bibitem[\protect\citeauthoryear{{Fragkoudi}, {Grand}, {Pakmor}, {G{\'o}mez},
  {Marinacci}  \& {Springel}}{{Fragkoudi} et~al.}{2025}]{Fragkoudi2024}
{Fragkoudi} F.,  {Grand} R. J.~J.,  {Pakmor} R.,  {G{\'o}mez} F.,  {Marinacci}
  F.,   {Springel} V.,  2025, \mn@doi [\mnras] {10.1093/mnras/staf389}, \href
  {https://ui.adsabs.harvard.edu/abs/2025MNRAS.538.1587F} {538, 1587}

\bibitem[\protect\citeauthoryear{{Gargiulo} et~al.,}{{Gargiulo}
  et~al.}{2019}]{Gargiulo2019}
{Gargiulo} I.~D.,  et~al., 2019, \mn@doi [\mnras] {10.1093/mnras/stz2536},
  \href {https://ui.adsabs.harvard.edu/abs/2019MNRAS.489.5742G} {489, 5742}

\bibitem[\protect\citeauthoryear{{Genel} et~al.,}{{Genel}
  et~al.}{2019}]{Genel2019}
{Genel} S.,  et~al., 2019, \mn@doi [\apj] {10.3847/1538-4357/aaf4bb}, \href
  {https://ui.adsabs.harvard.edu/abs/2019ApJ...871...21G} {871, 21}

\bibitem[\protect\citeauthoryear{{G{\'o}mez}, {White}, {Marinacci}, {Slater},
  {Grand}, {Springel}  \& {Pakmor}}{{G{\'o}mez} et~al.}{2016}]{Gomez2016}
{G{\'o}mez} F.~A.,  {White} S.~D.~M.,  {Marinacci} F.,  {Slater} C.~T.,
  {Grand} R.~J.~J.,  {Springel} V.,   {Pakmor} R.,  2016, \mn@doi [\mnras]
  {10.1093/mnras/stv2786}, \href
  {http://adsabs.harvard.edu/abs/2016MNRAS.456.2779G} {456, 2779}

\bibitem[\protect\citeauthoryear{{G{\'o}mez}, {White}, {Grand}, {Marinacci},
  {Springel}  \& {Pakmor}}{{G{\'o}mez} et~al.}{2017}]{Gomez2016b}
{G{\'o}mez} F.~A.,  {White} S.~D.~M.,  {Grand} R.~J.~J.,  {Marinacci} F.,
  {Springel} V.,   {Pakmor} R.,  2017, \mn@doi [\mnras]
  {10.1093/mnras/stw2957}, \href
  {http://adsabs.harvard.edu/abs/2017MNRAS.465.3446G} {465, 3446}

\bibitem[\protect\citeauthoryear{{Grand} et~al.,}{{Grand}
  et~al.}{2017}]{Auriga}
{Grand} R.~J.~J.,  et~al., 2017, \mn@doi [\mnras] {10.1093/mnras/stx071}, \href
  {http://adsabs.harvard.edu/abs/2017MNRAS.467..179G} {467, 179}

\bibitem[\protect\citeauthoryear{{Grand} et~al.,}{{Grand}
  et~al.}{2021}]{Grand2021}
{Grand} R. J.~J.,  et~al., 2021, \mn@doi [\mnras] {10.1093/mnras/stab2492},
  \href {https://ui.adsabs.harvard.edu/abs/2021MNRAS.507.4953G} {507, 4953}

\bibitem[\protect\citeauthoryear{{Hopkins} et~al.,}{{Hopkins}
  et~al.}{2020}]{Hopkins2019}
{Hopkins} P.~F.,  et~al., 2020, \mn@doi [\mnras] {10.1093/mnras/stz3321}, \href
  {https://ui.adsabs.harvard.edu/abs/2020MNRAS.492.3465H} {492, 3465}

\bibitem[\protect\citeauthoryear{{Irodotou} et~al.,}{{Irodotou}
  et~al.}{2022}]{Irodotou2022}
{Irodotou} D.,  et~al., 2022, \mn@doi [\mnras] {10.1093/mnras/stac1143}, \href
  {https://ui.adsabs.harvard.edu/abs/2022MNRAS.513.3768I} {513, 3768}

\bibitem[\protect\citeauthoryear{{Joshi}, {Pontzen}, {Agertz}, {Rey}, {Read}
  \& {Renaud}}{{Joshi} et~al.}{2024}]{Joshi2024}
{Joshi} G.~D.,  {Pontzen} A.,  {Agertz} O.,  {Rey} M.~P.,  {Read} J.,
  {Renaud} F.,  2024, \mn@doi [\mnras] {10.1093/mnras/stae129}, \href
  {https://ui.adsabs.harvard.edu/abs/2024MNRAS.528.2346J} {528, 2346}

\bibitem[\protect\citeauthoryear{{Keller}, {Wadsley}, {Wang}  \&
  {Kruijssen}}{{Keller} et~al.}{2019}]{Keller2019}
{Keller} B.~W.,  {Wadsley} J.~W.,  {Wang} L.,   {Kruijssen} J.~M.~D.,  2019,
  \mn@doi [\mnras] {10.1093/mnras/sty2859}, \href
  {https://ui.adsabs.harvard.edu/abs/2019MNRAS.482.2244K} {482, 2244}

\bibitem[\protect\citeauthoryear{{Marinacci}, {Pakmor}  \&
  {Springel}}{{Marinacci} et~al.}{2014}]{Marinacci2014}
{Marinacci} F.,  {Pakmor} R.,   {Springel} V.,  2014, \mn@doi [\mnras]
  {10.1093/mnras/stt2003}, \href
  {http://adsabs.harvard.edu/abs/2014MNRAS.437.1750M} {437, 1750}

\bibitem[\protect\citeauthoryear{{Martin-Alvarez}, {Sijacki}, {Haehnelt},
  {Farcy}, {Dubois}, {Belokurov}, {Rosdahl}  \&
  {Lopez-Rodriguez}}{{Martin-Alvarez} et~al.}{2023}]{MartinAlvarez2023b}
{Martin-Alvarez} S.,  {Sijacki} D.,  {Haehnelt} M.~G.,  {Farcy} M.,  {Dubois}
  Y.,  {Belokurov} V.,  {Rosdahl} J.,   {Lopez-Rodriguez} E.,  2023, \mn@doi
  [\mnras] {10.1093/mnras/stad2559}, \href
  {https://ui.adsabs.harvard.edu/abs/2023MNRAS.525.3806M} {525, 3806}

\bibitem[\protect\citeauthoryear{{Monachesi}, {G{\'o}mez}, {Grand},
  {Kauffmann}, {Marinacci}, {Pakmor}, {Springel}  \& {Frenk}}{{Monachesi}
  et~al.}{2016}]{Monachesi2016}
{Monachesi} A.,  {G{\'o}mez} F.~A.,  {Grand} R.~J.~J.,  {Kauffmann} G.,
  {Marinacci} F.,  {Pakmor} R.,  {Springel} V.,   {Frenk} C.~S.,  2016, \mn@doi
  [\mnras] {10.1093/mnrasl/slw052}, \href
  {http://adsabs.harvard.edu/abs/2016MNRAS.459L..46M} {459, L46}

\bibitem[\protect\citeauthoryear{{Monachesi} et~al.,}{{Monachesi}
  et~al.}{2019}]{Monachesi2019}
{Monachesi} A.,  et~al., 2019, \mn@doi [\mnras] {10.1093/mnras/stz538}, \href
  {https://ui.adsabs.harvard.edu/abs/2019MNRAS.485.2589M} {485, 2589}

\bibitem[\protect\citeauthoryear{{Nelson} et~al.,}{{Nelson}
  et~al.}{2019}]{Nelson2019TNGPublic}
{Nelson} D.,  et~al., 2019, \mn@doi [Computational Astrophysics and Cosmology]
  {10.1186/s40668-019-0028-x}, \href
  {https://ui.adsabs.harvard.edu/abs/2019ComAC...6....2N} {6, 2}

\bibitem[\protect\citeauthoryear{{Pakmor}, {Marinacci}  \& {Springel}}{{Pakmor}
  et~al.}{2014}]{Pakmor2014}
{Pakmor} R.,  {Marinacci} F.,   {Springel} V.,  2014, \mn@doi [\apjl]
  {10.1088/2041-8205/783/1/L20}, \href
  {http://adsabs.harvard.edu/abs/2014ApJ...783L..20P} {783, L20}

\bibitem[\protect\citeauthoryear{{Pakmor}, {Springel}, {Bauer}, {Mocz},
  {Munoz}, {Ohlmann}, {Schaal}  \& {Zhu}}{{Pakmor} et~al.}{2016}]{Pakmor2016}
{Pakmor} R.,  {Springel} V.,  {Bauer} A.,  {Mocz} P.,  {Munoz} D.~J.,
  {Ohlmann} S.~T.,  {Schaal} K.,   {Zhu} C.,  2016, \mn@doi [\mnras]
  {10.1093/mnras/stv2380}, \href
  {http://adsabs.harvard.edu/abs/2016MNRAS.455.1134P} {455, 1134}

\bibitem[\protect\citeauthoryear{{Pakmor} et~al.,}{{Pakmor}
  et~al.}{2017}]{Pakmor2017}
{Pakmor} R.,  et~al., 2017, \mn@doi [\mnras] {10.1093/mnras/stx1074}, \href
  {http://adsabs.harvard.edu/abs/2017MNRAS.469.3185P} {469, 3185}

\bibitem[\protect\citeauthoryear{{Pakmor}, {Guillet}, {Pfrommer}, {G{\'o}mez},
  {Grand}, {Marinacci}, {Simpson}  \& {Springel}}{{Pakmor}
  et~al.}{2018}]{Pakmor2018}
{Pakmor} R.,  {Guillet} T.,  {Pfrommer} C.,  {G{\'o}mez} F.~A.,  {Grand} R.
  J.~J.,  {Marinacci} F.,  {Simpson} C.~M.,   {Springel} V.,  2018, \mn@doi
  [\mnras] {10.1093/mnras/sty2601}, \href
  {https://ui.adsabs.harvard.edu/abs/2018MNRAS.481.4410P} {481, 4410}

\bibitem[\protect\citeauthoryear{{Pakmor} et~al.,}{{Pakmor}
  et~al.}{2020}]{Pakmor2020}
{Pakmor} R.,  et~al., 2020, \mn@doi [\mnras] {10.1093/mnras/staa2530}, \href
  {https://ui.adsabs.harvard.edu/abs/2020MNRAS.498.3125P} {498, 3125}

\bibitem[\protect\citeauthoryear{{Pakmor} et~al.,}{{Pakmor}
  et~al.}{2023}]{Pakmor2023MTNG}
{Pakmor} R.,  et~al., 2023, \mn@doi [\mnras] {10.1093/mnras/stac3620}, \href
  {https://ui.adsabs.harvard.edu/abs/2023MNRAS.524.2539P} {524, 2539}

\bibitem[\protect\citeauthoryear{{Pakmor} et~al.,}{{Pakmor}
  et~al.}{2024}]{Pakmor2024}
{Pakmor} R.,  et~al., 2024, \mn@doi [\mnras] {10.1093/mnras/stae112}, \href
  {https://ui.adsabs.harvard.edu/abs/2024MNRAS.528.2308P} {528, 2308}

\bibitem[\protect\citeauthoryear{{Pakmor} et~al.,}{{Pakmor}
  et~al.}{2025}]{Pakmor2025b}
{Pakmor} R.,  et~al., 2025, \mn@doi [arXiv e-prints]
  {10.48550/arXiv.2507.22104}, \href
  {https://ui.adsabs.harvard.edu/abs/2025arXiv250722104P} {p. arXiv:2507.22104}

\bibitem[\protect\citeauthoryear{{Pillepich} et~al.,}{{Pillepich}
  et~al.}{2018}]{Pillepich2018}
{Pillepich} A.,  et~al., 2018, \mn@doi [\mnras] {10.1093/mnras/stx3112}, \href
  {https://ui.adsabs.harvard.edu/abs/2018MNRAS.475..648P} {475, 648}

\bibitem[\protect\citeauthoryear{{Rey} et~al.,}{{Rey} et~al.}{2023}]{Rey2023}
{Rey} M.~P.,  et~al., 2023, \mn@doi [\mnras] {10.1093/mnras/stad513}, \href
  {https://ui.adsabs.harvard.edu/abs/2023MNRAS.521..995R} {521, 995}

\bibitem[\protect\citeauthoryear{{Riley} et~al.,}{{Riley}
  et~al.}{2025}]{Riley2024}
{Riley} A.~H.,  et~al., 2025, \mn@doi [\mnras] {10.1093/mnras/staf1350}, \href
  {https://ui.adsabs.harvard.edu/abs/2025MNRAS.542.2443R} {542, 2443}

\bibitem[\protect\citeauthoryear{{Rodr{\'\i}guez Montero}, {Martin-Alvarez},
  {Slyz}, {Devriendt}, {Dubois}  \& {Sijacki}}{{Rodr{\'\i}guez Montero}
  et~al.}{2024}]{Montero2024}
{Rodr{\'\i}guez Montero} F.,  {Martin-Alvarez} S.,  {Slyz} A.,  {Devriendt} J.,
   {Dubois} Y.,   {Sijacki} D.,  2024, \mn@doi [\mnras]
  {10.1093/mnras/stae1083}, \href
  {https://ui.adsabs.harvard.edu/abs/2024MNRAS.530.3617R} {530, 3617}

\bibitem[\protect\citeauthoryear{{Schaye} et~al.,}{{Schaye}
  et~al.}{2015}]{Schaye2005Eagle}
{Schaye} J.,  et~al., 2015, \mn@doi [\mnras] {10.1093/mnras/stu2058}, \href
  {http://adsabs.harvard.edu/abs/2015MNRAS.446..521S} {446, 521}

\bibitem[\protect\citeauthoryear{{Sellwood} \& {Debattista}}{{Sellwood} \&
  {Debattista}}{2009}]{SellwoodDebattista2009}
{Sellwood} J.~A.,  {Debattista} V.~P.,  2009, \mn@doi [\mnras]
  {10.1111/j.1365-2966.2009.15219.x}, \href
  {https://ui.adsabs.harvard.edu/abs/2009MNRAS.398.1279S} {398, 1279}

\bibitem[\protect\citeauthoryear{{Shipp} et~al.,}{{Shipp}
  et~al.}{2025}]{Shipp2024}
{Shipp} N.,  et~al., 2025, \mn@doi [\mnras] {10.1093/mnras/staf1283}, \href
  {https://ui.adsabs.harvard.edu/abs/2025MNRAS.542.1109S} {542, 1109}

\bibitem[\protect\citeauthoryear{{Simpson}, {Grand}, {G{\'o}mez}, {Marinacci},
  {Pakmor}, {Springel}, {Campbell}  \& {Frenk}}{{Simpson}
  et~al.}{2018}]{Simpson2018}
{Simpson} C.~M.,  {Grand} R. J.~J.,  {G{\'o}mez} F.~A.,  {Marinacci} F.,
  {Pakmor} R.,  {Springel} V.,  {Campbell} D. J.~R.,   {Frenk} C.~S.,  2018,
  \mn@doi [\mnras] {10.1093/mnras/sty774}, \href
  {https://ui.adsabs.harvard.edu/abs/2018MNRAS.478..548S} {478, 548}

\bibitem[\protect\citeauthoryear{{Simpson} et~al.,}{{Simpson}
  et~al.}{2019}]{Simpson2019}
{Simpson} C.~M.,  et~al., 2019, \mn@doi [\mnras] {10.1093/mnrasl/slz142}, \href
  {https://ui.adsabs.harvard.edu/abs/2019MNRAS.490L..32S} {490, L32}

\bibitem[\protect\citeauthoryear{{Springel}}{{Springel}}{2005}]{Springel2005b}
{Springel} V.,  2005, \mn@doi [MNRAS] {10.1111/j.1365-2966.2005.09655.x}, \href
  {http://adsabs.harvard.edu/abs/2005MNRAS.364.1105S} {364, 1105}

\bibitem[\protect\citeauthoryear{{Springel}}{{Springel}}{2010}]{Arepo}
{Springel} V.,  2010, \mn@doi [\mnras] {10.1111/j.1365-2966.2009.15715.x},
  \href {http://adsabs.harvard.edu/abs/2010MNRAS.401..791S} {401, 791}

\bibitem[\protect\citeauthoryear{{Springel} \& {Hernquist}}{{Springel} \&
  {Hernquist}}{2003}]{Springel2003}
{Springel} V.,  {Hernquist} L.,  2003, \mn@doi [\mnras]
  {10.1046/j.1365-8711.2003.06206.x}, \href
  {http://adsabs.harvard.edu/abs/2003MNRAS.339..289S} {339, 289}

\bibitem[\protect\citeauthoryear{{Springel} et~al.,}{{Springel}
  et~al.}{2005}]{Millenium}
{Springel} V.,  et~al., 2005, \mn@doi [Nature] {10.1038/nature03597}, \href
  {http://adsabs.harvard.edu/abs/2005Natur.435..629S} {435, 629}

\bibitem[\protect\citeauthoryear{{Springel}, {Frenk}  \& {White}}{{Springel}
  et~al.}{2006}]{Springel2006}
{Springel} V.,  {Frenk} C.~S.,   {White} S.~D.~M.,  2006, \mn@doi [\nat]
  {10.1038/nature04805}, \href
  {http://adsabs.harvard.edu/abs/2006Natur.440.1137S} {440, 1137}

\bibitem[\protect\citeauthoryear{{Springel} et~al.,}{{Springel}
  et~al.}{2008}]{Aquarius2}
{Springel} V.,  et~al., 2008, \mn@doi [MNRAS]
  {10.1111/j.1365-2966.2008.14066.x}, \href
  {http://adsabs.harvard.edu/abs/2008MNRAS.391.1685S} {391, 1685}

\bibitem[\protect\citeauthoryear{{Springel} et~al.,}{{Springel}
  et~al.}{2018}]{Springel2018}
{Springel} V.,  et~al., 2018, \mn@doi [\mnras] {10.1093/mnras/stx3304}, \href
  {https://ui.adsabs.harvard.edu/abs/2018MNRAS.475..676S} {475, 676}

\bibitem[\protect\citeauthoryear{{Springel}, {Pakmor}, {Zier}  \&
  {Reinecke}}{{Springel} et~al.}{2021}]{Gadget4}
{Springel} V.,  {Pakmor} R.,  {Zier} O.,   {Reinecke} M.,  2021, \mn@doi
  [\mnras] {10.1093/mnras/stab1855}, \href
  {https://ui.adsabs.harvard.edu/abs/2021MNRAS.506.2871S} {506, 2871}

\bibitem[\protect\citeauthoryear{{Vera-Casanova.} et~al.,}{{Vera-Casanova.}
  et~al.}{2025}]{Vera2025}
{Vera-Casanova.} A.,  et~al., 2025, \mn@doi [arXiv e-prints]
  {10.48550/arXiv.2503.17202}, \href
  {https://ui.adsabs.harvard.edu/abs/2025arXiv250317202V} {p. arXiv:2503.17202}

\bibitem[\protect\citeauthoryear{{Villaescusa-Navarro}
  et~al.,}{{Villaescusa-Navarro} et~al.}{2021}]{VillaescusaNavarro2021}
{Villaescusa-Navarro} F.,  et~al., 2021, \mn@doi [\apj]
  {10.3847/1538-4357/abf7ba}, \href
  {https://ui.adsabs.harvard.edu/abs/2021ApJ...915...71V} {915, 71}

\bibitem[\protect\citeauthoryear{{Vogelsberger}, {Genel}, {Sijacki}, {Torrey},
  {Springel}  \& {Hernquist}}{{Vogelsberger} et~al.}{2013}]{Vogelsberger2013}
{Vogelsberger} M.,  {Genel} S.,  {Sijacki} D.,  {Torrey} P.,  {Springel} V.,
  {Hernquist} L.,  2013, \mn@doi [\mnras] {10.1093/mnras/stt1789}, \href
  {http://adsabs.harvard.edu/abs/2013MNRAS.436.3031V} {436, 3031}

\bibitem[\protect\citeauthoryear{{Vogelsberger} et~al.,}{{Vogelsberger}
  et~al.}{2014}]{Vogelsberger2014Illustris}
{Vogelsberger} M.,  et~al., 2014, \mn@doi [\nat] {10.1038/nature13316}, \href
  {http://adsabs.harvard.edu/abs/2014Natur.509..177V} {509, 177}

\bibitem[\protect\citeauthoryear{{Vogelsberger}, {Marinacci}, {Torrey}  \&
  {Puchwein}}{{Vogelsberger} et~al.}{2020}]{Vogelsberger2020}
{Vogelsberger} M.,  {Marinacci} F.,  {Torrey} P.,   {Puchwein} E.,  2020,
  \mn@doi [Nature Reviews Physics] {10.1038/s42254-019-0127-2}, \href
  {https://ui.adsabs.harvard.edu/abs/2020NatRP...2...42V} {2, 42}

\bibitem[\protect\citeauthoryear{{Weinberger}, {Springel}  \&
  {Pakmor}}{{Weinberger} et~al.}{2020}]{Weinberger2020}
{Weinberger} R.,  {Springel} V.,   {Pakmor} R.,  2020, \mn@doi [\apjs]
  {10.3847/1538-4365/ab908c}, \href
  {https://ui.adsabs.harvard.edu/abs/2020ApJS..248...32W} {248, 32}

\bibitem[\protect\citeauthoryear{{Whittingham}, {Sparre}, {Pfrommer}  \&
  {Pakmor}}{{Whittingham} et~al.}{2021}]{Whittingham2021}
{Whittingham} J.,  {Sparre} M.,  {Pfrommer} C.,   {Pakmor} R.,  2021, \mn@doi
  [\mnras] {10.1093/mnras/stab1425}, \href
  {https://ui.adsabs.harvard.edu/abs/2021MNRAS.506..229W} {506, 229}

\bibitem[\protect\citeauthoryear{{Whittingham}, {Sparre}, {Pfrommer}  \&
  {Pakmor}}{{Whittingham} et~al.}{2023}]{Whittingham2023}
{Whittingham} J.,  {Sparre} M.,  {Pfrommer} C.,   {Pakmor} R.,  2023, \mn@doi
  [\mnras] {10.1093/mnras/stad2680}, \href
  {https://ui.adsabs.harvard.edu/abs/2023MNRAS.526..224W} {526, 224}

\bibitem[\protect\citeauthoryear{{van de Voort}, {Bieri}, {Pakmor},
  {G{\'o}mez}, {Grand}  \& {Marinacci}}{{van de Voort}
  et~al.}{2021}]{vandeVoort2021}
{van de Voort} F.,  {Bieri} R.,  {Pakmor} R.,  {G{\'o}mez} F.~A.,  {Grand} R.
  J.~J.,   {Marinacci} F.,  2021, \mn@doi [\mnras] {10.1093/mnras/staa3938},
  \href {https://ui.adsabs.harvard.edu/abs/2021MNRAS.501.4888V} {501, 4888}

\makeatother
\end{thebibliography}





\bsp	
\label{lastpage}
\end{document}